\documentclass[prd,draft,eqsecnum,showpacs,aps,nofootinbib]{revtex4}
\usepackage{latexsym}
\usepackage[dvips]{graphicx} 

\newcommand{\beq}{\begin{equation}}
\newcommand{\eeq}{\end{equation}}
\newcommand{\beqarr}{\begin{eqnarray}}
\newcommand{\eeqarr}{\end{eqnarray}}

\begin{document}
\draft

%
%
%
\renewcommand{\topfraction}{0.99}
\renewcommand{\bottomfraction}{0.99}
\title{The Full Second-Order Radiation Transfer Function for 
Large-Scale CMB Anisotropies}

\author{Nicola Bartolo}\email{nbartolo@ictp.trieste.it}
\affiliation{The Abdus Salam International Centre for Theoretical 
Physics, Strada Costiera 11, 34100 Trieste, Italy}

\author{Sabino Matarrese}\email{sabino.matarrese@pd.infn.it}
\affiliation{Dipartimento di Fisica ``G.\ Galilei'', Universit\`{a} di Padova, 
        and INFN, Sezione di Padova, via Marzolo 8, Padova I-35131, Italy}

\author{Antonio Riotto}\email{antonio.riotto@pd.infn.it}
\affiliation{CERN, Theory Division, CH-1211 Geneva 23, Switzerland}

\date{\today}
\vskip 2cm
\begin{abstract}
\noindent 
We calculate the full second-order radiation transfer function for 
Cosmic Microwave Background anisotropies on large angular scales 
in a flat universe filled with matter and cosmological constant. 
It includes $(i)$ the 
second-order generalization of the Sachs-Wolfe effect, and of $(ii)$ both the early and late Integrated Sachs-Wolfe 
effects, $(iii)$ the contribution of the second-order tensor modes, 
and is valid for a generic set of initial conditions specifying the level of primordial non-Gaussianity. 
\end{abstract}

\pacs{98.80.Cq \hfill DFPD--A/05/23}
\maketitle
\vskip2pc
\section{Introduction}
 
Cosmological inflation \cite{lrreview} has become the dominant paradigm to 
understand the initial conditions for Cosmic Microwave Background (CMB)
anisotropies and structure formation. In the inflationary picture, 
the primordial cosmological perturbations are created from quantum 
fluctuations ``redshifted'' out of the horizon during an early period of 
accelerated expansion of the universe, where they remain ``frozen''.  
They are observable as temperature anisotropies in the CMB. 
They were first detected by the COsmic Background 
Explorer (COBE) satellite \cite{smoot92,bennett96,gorski96}and then 
with spectacular accuracy most recently by the the Wilkinson Microwave 
Anisotropy Probe (WMAP) mission~\cite{wmap1,wmap2}. 
Since the observed cosmological perturbations are of the order
of $10^{-5}$, one might think that first-order perturbation theory
will be adequate for all comparison with observations. That may not be
the case however, as the Planck satellite \cite{planck} and its
successors may be sensitive to the non-Gaussianity of the cosmological
perturbations at the level of second- or higher-order perturbation theory
\cite{review}. Statistics like the bispectrum and the trispectrum of the 
CMB can then be used to assess the level of primordial non-Gaussianity on 
various cosmological scales and to discriminate it from the one induced by 
secondary anisotropies and systematic effects \cite{review,hu,dt,jul}. 
Therefore, it is of fundamental importance to provide accurate
theoretical predictions for the statistics of CMB anisotropies as left 
imprinted by the primordial seeds originated during or
immediately after inflation. Steps towards this goal have been 
taken in Refs. \cite{pyne,mm,prl,tomita1,tomita2,tomita3} 
at the level of second order perturbation theory. 

In this paper we provide the first computation of the {\it complete}  
second-order radiation transfer function for CMB anisotropies on large 
angular scales in a flat $\Lambda$ Cold Dark Matter ($\Lambda$CDM) model. 
Our calculation includes the second-order generalization of the 
Sachs-Wolfe as well as of the (both early and late) Integrated Sachs-Wolfe 
effects and is valid for a generic set of initial conditions 
specifying the level of primordial non-Gaussianity. It also accounts for the contribution of second-order tensor 
perturbations. Our results which are complete as fas as large angular scales 
are concerned, represent the first step towards the 
computation of the full second-order radiation transfer function for 
CMB anisotropies. 
Its computation also on small angular scales, which involves several   
non-linear effects, such as gravitational lensing, Shapiro time-delay, 
Rees-Sciama effects, etc. and, 
of course, dealing with the non-linear Boltzmann equation for the 
photon-matter fluid (some specific aspects of 
which have been studied in Ref.~\cite{boltz}), will be the subject of a 
future publication \cite{fut}.  

The general expression for CMB anisotropies up to second order from 
gravitational perturbations has been obtained in Refs.~\cite{pyne,mm}. 
At large angular scales the extension up to second order of the Sachs-Wolfe 
effect \cite{sw} has been derived in the Poisson gauge 
(for adiabatic perturbations) in Refs.~\cite{JCAP,review} (and for a 
gauge-invariant generalization at second-order see Ref.~\cite{prl}) 
\footnote{For a fully non-linear generalization of the Sachs-Wolfe effect 
see Ref.~\cite{NP}.} 
\begin{equation}
\label{SW}
{\frac{\Delta T}{T}}=\frac{1}{3} \phi_* -\frac{5}{18} \phi^2_{1*}\, ,
\end{equation}
where $\phi_*=\phi_{1*}+\phi_{2*} /2 \equiv \varphi_* + \phi_{2*} /2$ 
is the gravitational potential at emission on the last scattering surface and 
is split into linear and second-order parts 
(see Eq.~(\ref{metric})). Such an expression allows to single out the 
initial (at last scattering) non-linearities of the large-scale  
CMB anisotropies, including the primordial non-Gaussianity of cosmological 
perturbations, as it has been computed in Refs.~\cite{JCAP,prl,review,NP}. 
In Eq.~(\ref{SW}) the large-scale 
second-order gravitational potential $\phi_{2*}$ at emission contains 
the non-linearity arising from inflation and from the post-inflationary epoch 
yielding~~\cite{JCAP,prl,review,NP}
\begin{eqnarray}
\label{SWa}
\frac{\Delta T}{T}=\frac{1}{3} \varphi_{*}+ 
\frac{1}{18}\varphi^2_{*}-\frac{{\cal K}}{10}-\frac{5}{9}
(a_{\rm nl}-1) \varphi^2_{*}\, ,
\end{eqnarray}     
where
\begin{equation}
\label{calK}
{\cal K} \equiv 10 \nabla^{-4} \partial_i \partial^j( \partial^i \varphi_{*} 
\partial_j \varphi_{*})
-\nabla^{-2} \left(\frac{10}{3} \partial^i \varphi_{*} \partial_i \varphi_{*} 
\right)\, ,
\end{equation}
and $a_{\rm nl}$ is a parameter specifying the level of primordial 
non-Gaussianity which depends upon the particular scenario for the 
generation of cosmological perturbations, as we will discuss later 
in more detail. For example, for standard single-field slow-roll models of 
inflation $a_{\rm nl}=1 + {\cal O}(\epsilon,\eta) 
$~\cite{ABMR,maldacena,BMR2}, where $\epsilon$ and $\eta$ are 
the standard slow-roll parameters (see {\it e.g.} Ref.~\cite{lrreview}). 

In this work we will focus on another contribution to the large-scale CMB 
anisotropies which is due to the evolution of the gravitational potentials 
from the epoch of last scattering up to now and it is encoded in a number of 
integrated terms  
~\cite{mm}
\begin{equation}
\label{main0}
\frac{\Delta T}{T}(\hat{{\bf n}})=\int_{\eta_*}^{\eta_0} d\eta\, 
\frac{\partial}{\partial \eta}(\phi+\psi)({\bf x},\eta) 
\Big|_{{\bf x}=-\hat{{\bf n}}(\eta_0-\eta)}+({\rm first-order})^2\, ,
\end{equation}       
where $\phi=\phi_1+\phi_2/2$ and $\psi=\psi_1+\psi_2 /2$. 
In Eq.~(\ref{main0}) $\hat{{\bf n}}$ denotes the direction of photon momentum
and $\eta$ is the conformal time, $\eta_0$ and $\eta_*$ being the present 
and the last scattering times, respectively. Note that the partial 
time derivative of the perturbations must be evaluated along the line of 
sight. These integrated terms represent the gravitational redshift due to 
the time variation of the metric during the photon travel from the 
last scattering surface to the observer. 
The linear contribution is the so-called Integrated Sachs-Wolfe (ISW)
effect and in a flat matter dominated (Einstein-de Sitter) Universe 
it vanishes due to the constancy in time of the gravitational potentials. 
However, in the presence of a cosmological constant 
(or more general dark energy component) it actually gives a large-scale 
contribution due to the decay in time of the potentials: this is the 
so-called {\it late} ISW effect. The {\it early} ISW effect similarly arises 
because of the residual dynamical role of the radiation component at times 
close to the last-scattering epoch; this also makes the gravitational 
potentials evolve in time, leading to a non-negligible contribution to 
the large-scale anisotropy (see {\it e.g.} Ref.~\cite{husu95}). 
The second-order generalization of the ISW effect is 
characterized by some peculiar properties. Even in a flat matter-dominated 
Universe, when the linear counterpart vanishes because the gravitational
potential is constant, the evolution of the perturbations in the non-linear 
regime generates a non-vanishing contribution to the CMB anisotropies, the 
so-called Rees-Sciama effect \cite{RS}.
In this paper we will compute the second-order ISW generated in a 
universe with a non-vanishing cosmological constant term. Unlike the usual Rees-Sciama effect, a contribution to 
the CMB anisotropies on large scales is generated, 
and therefore for the rest of the paper by second-order ISW we will 
refer to the large scale part of Eq.~(\ref{main0}). 
Such an effect has not been considered so far, and thus a major part of this 
paper will deal with its study. 
An important feature is that it is related to possible primordial 
non-Gaussianity in the energy-density perturbations produced in the early 
universe. In a separate paper we will give a quantitative study of its  
signatures in the bispectrum of the CMB anisotropies~\cite{fut}. 
Notice that in Eq.~(\ref{main0}) we have also indicated additional 
contributions to CMB anisotropies which arise from integrated terms 
quadratic in the linear metric perturbations~\cite{pyne,mm}. 
We will discuss and take them into account later. 
Finally in writing Eq.~(\ref{main0}) we have dropped possible 
gravitational redshifts from second-order vector modes 
which are expected to give a negligible contribution on large 
scales~\cite{mm}.

This paper serves for different purposes. First it contains an 
independent part about the evolution of the second-oder perturbations 
in a flat universe filled with matter and a cosmological constant term. 
Second it discusses the generation of CMB anisotropies due to the 
second-order ISW and its connection to primordial non-Gaussianity. 
The final achievement is that of providing the full second-order 
radiation transfer function for CMB anisotropies on large angular scales 
from the second-order Sachs-Wolfe and ISW effects. 
Our formalism is the starting point of an accurate evaluation of different 
effects on the CMB anisotropies, including the exact expression for the 
CMB bispectrum on large scales ~\cite{fut}. 

The paper is organized as follows. In Section II we present the main 
equations and solutions for the evolution of the gravitational potentials 
in a $\Lambda$CDM universe, accounting for primordial non-Gaussianity 
in the initial conditions (see also Appendix A).
In Section III we derive the expression for the second-order ISW by taking 
the large-scale contributions to Eq.~(\ref{main0}) and we derive an 
expression of the CMB anisotropies expanded in multipoles (with details 
in Appendix B). In Section IV we study the contribution to the CMB anisotropies from second-order tensor modes.
Finally in Section V we provide the full expression of the second-order radiation 
transfer function for large-scale CMB anisotropies. In this section we summarize our final formulae as a list 
and we provide the reader with all the necessary definitions.
Section V contains also our conclusions.    
 
\section{Second-order perturbations of a flat $\Lambda$CDM Universe in the 
Poisson Gauge}
\label{lambdasecond}

\subsection{Evolution of the second-order gravitational potentials}
We consider a spatially flat Universe filled with a cosmological constant 
$\Lambda$ and a non-relativistic pressureless fluid of Cold Dark Matter (CDM) 
plus baryons, whose energy-momentum tensor reads 
$T^{\mu}_{~\nu}=\rho u^{\mu} u_{\nu}$. The radiation component, which also
plays a role during the early stages following the last-scattering epoch, 
will be considered in Section IIIC, in connection with the early ISW effect. 
We will work in the so-called 
Poisson gauge which, to linear order, reduces to the Newtonian gauge. 
Using the conformal time $d \eta =dt/a$, the perturbed line element around 
a spatially flat FRW background reads
\begin{equation}
\label{metric}
ds^2=a^2(\eta)\{-(1+2\phi)d\eta^2 + 2 \omega_i d\eta dx^i+[(1-2\psi) 
\delta_{ij} + \chi_{ij}]dx^i dx^j \}\, .
\end{equation}
Here each perturbation quantity is expanded into a 
first-order (linear) part and a second-order contribution in the same
 way as for the 
gravitational potential $\phi=\phi_1+\phi_2/2$. 
In the Poisson gauge the shift perturbation 
$\omega_i$ is a pure vector, {\it i.e.} $\partial_i \omega^i=0$, while 
$\chi_{ij}$ is a tensor mode ({\it i.e.} divergence-free and traceless     
$\partial^i \chi_{ij}=0$, $\chi^i_{~i}=0$). In this paper we are mainly 
interested in the evolution of the second-order scalar potentials. Moreover 
we will neglect linear vector modes since they are not produced in standard 
mechanisms for the generation of cosmological perturbations (as inflation), 
and we also neglect tensor modes at linear order, since they give a negligible 
contribution to the bispectrum arising from the integrated 
effects~(\ref{main0}). In Section~\ref{Tensor2} and Appendix A, we will report also the computation 
for the second-order vector 
and tensor modes following the results of Ref.~\cite{mhm}. 
As for the matter component we split the mass density into a homogeneous 
$\bar{\rho}(\eta)$ and a perturbed part as 
$\rho({\bf x},\eta)=\bar{\rho}(\eta) (1+\delta_1+\delta_2/2)$ 
and we write the four velocity as 
$u^{\mu}=(\delta^{\mu}_0+v^{\mu})/a$ with $u^{\mu}u_{\mu}=-1$ and 
$v^\mu=v^\mu_1+v^\mu_2/2$.

The Friedmann background equations are $3{\mathcal H}^2
=a^2(8 \pi G {\bar \rho}+\Lambda)$ and 
${\bar \rho}'=-3{\mathcal H} {\bar \rho}$, 
where a prime stands for differentiation with 
respect to conformal time, and ${\mathcal H}=a'/a$.

Let us briefly recall the results for the linear perturbations in 
the case of a non-vanishing cosmological $\Lambda$ term. 
At linear order the traceless part of the ($i$-$j$)-components of
Einstein equations gives 
$\phi_1=\psi_1 \equiv \varphi$. Its trace gives the evolution 
equation for the linear 
scalar potential $\varphi$
\begin{equation}
\label{ev}
\varphi''+3 {\mathcal H} \varphi'+a^2 \Lambda \varphi=0\, .
\end{equation}
Selecting only the growing mode solution one can write
\begin{equation}
\label{relphiphi_0}
\varphi({\bf x}, \eta)= g(\eta)\, \varphi_0({\bf x}) \, ,
\end{equation}
where $\varphi_0$ is the peculiar gravitational potential linearly 
extrapolated to the 
present time $(\eta_0)$ and $g(\eta)=D_+(\eta)/a(\eta)$ is the so called 
growth-suppression factor, where $D_+(\eta)$ is the linear growing-mode 
of density 
fluctuations in the Newtonian limit. The exact form of $g$ can be found in 
Refs.~\cite{lahav,Carroll,Eisenstein}. 
In the $\Lambda=0$ case $g=1$. A very good approximation for $g$ as a 
function of redshift $z$ is given in 
Refs.~\cite{lahav,Carroll} 
\begin{equation}
g \propto \Omega_m\left[\Omega_m^{4/7} - \Omega_\Lambda +
\left(1+ \Omega_m/2\right)\left(1+ \Omega_\Lambda/70\right)\right]^{-1} \;, 
\end{equation}
with $\Omega_m=\Omega_{0m}(1+z)^3/E^2(z)$, 
$\Omega_\Lambda=\Omega_{0\Lambda}/E^2(z)$, 
$E(z) \equiv (1+z) {\mathcal H}(z)/{\mathcal H}_0 = \left[\Omega_{0m}(1+z)^3 + 
\Omega_{0\Lambda}\right]^{1/2}$ and 
$\Omega_{0m}$, $\Omega_{0\Lambda}=1-\Omega_{0m}$, the present-day
density parameters of non-relativistic matter and cosmological constant, 
respectively. We will normalize the growth-suppression factor so that 
$g(z=0)=1$. 

Let us now consider the second-order scalar perturbations. For simplicity,  
we adopt the notation
\begin{eqnarray}
\phi_2 &\equiv& \Phi \, ,\\
\psi_2 &\equiv&  \Psi\, .
\end{eqnarray}
The evolution equation for the second-order gravitational potential $\Psi$ 
is obtained from the trace of the ($i$-$j$)-Einstein equations:   
its detailed derivation will be given in Appendix~\ref{A}; 
here we just report it 
\begin{equation}
\label{Psieq}
\Psi''+3 {\mathcal H} \Psi'+a^2 \Lambda \Psi= S(\eta)\, ,
\end{equation}
where $S(\eta)$ is the source term   
\begin{eqnarray}
\label{PSIsource}
S(\eta)&=&g^2\Omega_m {\mathcal H}^2 \Bigg[\frac{(f-1)^2}{\Omega_m} 
\varphi_0^2+2 \Bigg(2 \frac{(f-1)^2}{\Omega_m} -\frac{3}{\Omega_m}+3 \Bigg) 
\times \Bigg( 
\nabla^{-2} 
\left( \partial^i \varphi_0 
\partial_i \varphi_0 \right)- 3 \nabla^{-4} \partial_i \partial^j 
\left(\partial^i \varphi_0 \partial_j \varphi_0 \right) \Bigg)
\Bigg] \nonumber \\
&+& g^2 \Bigg[ \frac{4}{3}   
\left( \frac{f^2}{\Omega_m}+\frac{3}{2} \right) 
\nabla^{-2}\partial_i \partial^j
\left(\partial^i \varphi_0 \partial_j \varphi_0\right) - 
\left( \partial^i \varphi_0 
\partial_i \varphi_0 \right) \Bigg] \, ,
\end{eqnarray}
and we have introduced the function 
\begin{equation}
f(\eta)=\frac{d \ln D_+}{d \ln a}=1+\frac{g'(\eta)}{{\mathcal H} g(\eta)}\, ,
\end{equation}
which can be written as a function of $\Omega_m$ as $f(\Omega_m) 
\approx \Omega_m(z)^{4/7}$
~\cite{lahav,Carroll}. In Eq.~(\ref{PSIsource}) $\nabla^{-2}$ stands 
for the inverse of the Laplacian operator. 

The solution of Eq.~(\ref{Psieq}) is then given by 
\begin{eqnarray}
\label{solution0}
\Psi(\eta)=\frac{g}{g_{m}} \Psi_{m}+\Psi_+(\eta) 
\int_{\eta_m}^\eta d\eta' \frac{\Psi_-(\eta')}{W(\eta')} S(\eta')-\Psi_-(\eta)
\int_{\eta_m}^\eta d\eta' \frac{\Psi_+(\eta')}{W(\eta')} S(\eta')\, .
\end{eqnarray} 
Here $\Psi_+(\eta)=g(\eta)$ and $\Psi_-(\eta)={\mathcal H}(\eta)/a^2(\eta)$ 
are the solutions of 
the homogeneous equation, $W$ is the Wronskian which explicitly reads 
$W(\eta)=W_0/a^3$ with $W_0={\mathcal H}_0^2 (f_0+(3/2) \Omega_{0m} )$, where 
the suffix `0' stands for the value of the corresponding quantities at the 
present time. In particular notice that $\Psi_{m} \equiv \Psi(\eta_m)$ represents the 
initial condition taken deep in the matter dominated era on super-horizon 
scales, $\eta_m$ being the epoch when full matter domination starts. It is such an initial value 
that must be properly determined in order to account for the primordial 
non-Gaussianity in the cosmological perturbations. 
The evolution of the gravitational potential $\Phi$ is then obtained from the 
relation between $\Psi$ and $\Phi$
\begin{equation}
\label{rela}
\nabla^2 \nabla^2 \Psi=\nabla^2 \nabla^2 \Phi -4g^2 \nabla^2 \nabla^2 
\varphi^2_0 -\frac{4}{3} g^2 \Big( \frac{f^2}{\Omega_m}+\frac{3}{2} \Big) 
\Big[ \nabla^2 (\partial_i \varphi_0 \partial^i \varphi_0) -3 \partial_i 
\partial^j (\partial^i\varphi_0 \partial_j \varphi_0) \Big]\, ,
\end{equation}
which follows from the traceless part of the ($i$-$j$)-component of 
Einstein equations 
(see Appendix~\ref{A}). 
In particular deep in the matter dominated epoch Eq.~(\ref{rela}) reduces to 
\begin{equation}
\label{relain}
\nabla^2 \nabla^2 \Psi_{m}=\nabla^2 \nabla^2 \Phi_{m}-
4 g_{m}^2 \nabla^2 \nabla^2 \varphi_0^2-\frac{10}{3} 
g^2_{m}\left[\nabla^2 (\partial^i \varphi_0 \partial_i \varphi_0 )-
3 \partial^i\partial_j (\partial_i\varphi_0\partial^j\varphi_0)\right]\, ,
\end{equation}  
where we have used that for $\Omega_m \rightarrow 1$ $f(\eta) 
\rightarrow 1$, and 
where $\varphi_m=g_{m} \varphi_0$ is the value of the 
gravitational 
potential during matter domination, when the cosmological constant was still 
negligible, with $g_{m}=g(\eta_m)$.\footnote{A good approximation, $g_m 
\approx \frac{2}{5} \Omega_{0m}^{-1}(\Omega_{0m}^{4/7} + 
\frac{3}{2}\Omega_{0m})$.} 
Note that in this way Eq.~(\ref{relain}) reduces to the relation 
for the two gravitational potentials 
$\Phi$ and $\Psi$ first obtained in Ref.~\cite{BMR2} for a 
matter-dominated Universe.

\subsection{Initial conditions from primordial non-Gaussianity}
Let us now discuss the key issue of the  initial 
conditions which we conveniently 
fix at the   time when the cosmological perturbations
relevant today for Large Scale Structures and CMB anisotropies are well outside the Hubble radius, 
{\emph i.e.} when the (comoving) wavelength $\lambda$ 
of a given perturbation mode is such that
 $\lambda \gg {\mathcal H}^{-1}$,  
${\mathcal H}=a'/a$ being the horizon size. 

In the standard single-field inflationary model, 
the first seeds of density fluctuations are generated on super-horizon scales 
from the fluctuations of a scalar field, the inflaton \cite{lrreview}. 
Recently many other scenarios have been proposed as alternative 
mechanisms to generate such primordial seeds. 
These include, for example, the curvaton~\cite{curvaton} and the 
inhomogeneous reheating scenarios~\cite{gamma1}, 
where essentially the first density 
fluctuations are produced through the fluctuations of a scalar field
other than the inflaton. 
In order to follow the evolution on super-horizon scales of the  
density fluctuations coming from the various  mechanisms, we 
use the curvature perturbation of uniform density hypersurfaces $\zeta=\zeta_1+\zeta_2/2+\cdots$, where 
$\zeta_1=-\psi_1-{\mathcal H} {\delta \rho_1}/{\bar{\rho}'}$ and
the expression for $ \zeta_2$ is given by~\cite{mw}
\begin{equation}
\label{defz2}
\zeta_2=-\psi_2-{\cal H} \frac{\delta_2 \rho}{{\rho}'}+\Delta \zeta_2\, ,
\end{equation}
with
\begin{equation}
\label{deltaz2}
\Delta \zeta_2 = 2 {\cal H} \frac{\delta_1 \rho'}{{\rho}'} \frac{\delta_1 \rho}{\rho'}+2
\frac{\delta_1 \rho}{\rho'} (\psi_1'+2{\cal H} \psi_1) - 
\left( \frac{\delta_1 \rho}{\rho'} \right)^2 \left({\cal H} \frac{\rho''}{\rho} -{\cal H}' 
-2{\cal H}^2  \right)\, .
\end{equation}
The crucial point is that the  gauge-invariant curvature perturbation
$\zeta$ remains  {\it constant} on super-horizon scales after it 
has been generated during a primordial epoch and possible isocurvature 
perturbations are no longer present. Therefore, we may set
the initial conditions at the time when $\zeta$  becomes
constant. In particular,  $\zeta_2$ 
provides  the necessary information about the
``primordial'' level of non-Gaussianity generated either during inflation, 
as in the 
standard scenario, or immediately after it, as in the curvaton scenario. 
Different scenarios are  characterized by different values of 
$\zeta_2$. For example, in    
the standard single-field inflationary model
 $\zeta_2=2\left( 
\zeta_1\right)^2+{\cal O}\left(\epsilon,\eta\right)$~\cite{ABMR,BMR2}, where 
$\epsilon$ and $\eta$ are the standard slow-roll parameters~\cite{lrreview}. 
In general, we may  parametrize the primordial non-Gaussianity level 
in terms of the conserved curvature perturbation as in Ref. \cite{prl}  
\begin{equation}
\label{param}
\zeta_2=2 a_{\rm nl}\left(\zeta_1\right)^2\, ,
\end{equation}
where the parameter $a_{\rm nl}$ depends on the physics of a given scenario.
For example in the standard scenario $a_{\rm nl}\simeq 1$, while in the  
curvaton case $a_{\rm nl}=(3/4r)-r/2$, where 
$r \approx (\rho_\sigma/\rho)_{\rm D}$ is the relative   
curvaton contribution to the total energy density at curvaton 
decay~\cite{ngcurv,review}. In the minimal picture for the inhomogeneous 
reheating scenario, $a_{\rm nl}=1/4$. For other scenarios we refer 
the reader to Ref.~\cite{review}. 
One of the best tools 
to detect or constrain the primordial large-scale non-Gaussianity is 
through the analysis 
of the CMB anisotropies, for example by studying the bispectrum
~\cite{review}. In that case the standard procedure is to
introduce  the non-linearity 
parameter $f_{\rm nl}$ characterizing non-Gaussianity in the large-scale 
temperature anisotropies~\cite{ks,k,review}. To give the feeling
of the resulting size of $f_{\rm nl}$ when $|a_{\rm nl}| \gg 1$, 
$f_{\rm nl} 
\simeq 5 a_{\rm nl}/3$~(see Refs.~\cite{review,prl}).
The conserved value of the curvature perturbation $\zeta$
allows to  set the initial 
conditions for the metric and matter perturbations accounting for the 
primordial contributions. At linear order during the matter-dominated epoch 
and on large scales 
$\zeta_1=-5 \varphi_{m}/3$, where 
$\varphi_{m}=g_{m} \varphi_0$ is the value of the gravitational 
potential during matter domination, when the cosmological constant was still 
negligible. Thus we can write   
\begin{equation}
\label{zetain}
\zeta_2=\frac{50}{9}\,a_{\rm nl}\, \varphi_{m}^2=
\frac{50}{9}\,a_{\rm nl}\, g_{m}^2 \varphi_{0}^2\, .
\end{equation} 
On the other hand by using the expression for $\zeta_2$ during the 
matter-dominated epoch together with 
Eq.~(\ref{relain}) and 
the second-order ($0$-$0$)-component of Einstein equations 
(evaluated for a matter-dominated epoch), one can express 
$\Phi_{m}$ in terms of $\zeta_2$ as 
(see Refs.~\cite{BMR2,JCAP,review,prl})    
\begin{equation}
\label{phiz}
\Phi_{m}=-\frac{3}{5} \zeta_2 +\frac{16}{3} \varphi_{m}^2+2 
\nabla^{-2}(\partial^i \varphi_{m} \partial_i \varphi_{m}) 
- 6 \nabla^{-4}\partial_i
\partial^j(\partial^i\varphi_{m} \partial_j \varphi_{m})\, .
\end{equation} 
Therefore using Eq.~(\ref{zetain}) we find, 
for generic non-Gaussian initial conditions parametrized by $\zeta_2=2 
a_{\rm nl} \left(\zeta_1 \right)^2$,  
\begin{eqnarray}
\Phi_{m}&=&2 g_{m}^2 \Bigg[
\left( -\frac{5}{3}(a_{\rm nl}-1)+1 \right) \varphi_0^2+ \nabla^{-2} 
\left( \partial^i \varphi_0 
\partial_i \varphi_0 \right)- 3 \nabla^{-4} \partial_i \partial^j
\left(\partial^i \varphi_0 \partial_j \varphi_0 \right) \Bigg] \, ,\\
\label{Psiinfty}
\Psi_{m}&=&2 g_{m}^2 \Bigg[
\left( -\frac{5}{3}(a_{\rm nl}-1)-1 \right) \varphi_0^2-\frac{2}{3} \left( 
\nabla^{-2} 
\left( \partial^i \varphi_0 
\partial_i \varphi_0 \right)- 3 \nabla^{-4} \partial_i \partial^j
\left(\partial^i \varphi_0 \partial_j \varphi_0 \right) \right)\Bigg] \, ,
\nonumber \\ 
\end{eqnarray}
where $a_{\rm nl}$ will  always signal the presence of primordial 
non-Gaussianity 
according to our parametrization. The last of these equations~(\ref{Psiinfty}) 
has been obtained from Eq.~(\ref{relain}).

After having determined the initial conditions from Eq.~(\ref{solution0}) 
we finally get
\begin{eqnarray}
\Psi(\eta) & = & 2 g(\eta)g_{m} 
\Bigg[
\left( -\frac{5}{3}(a_{\rm nl}-1)-1 \right) \varphi_0^2-\frac{2}{3} \left( 
\nabla^{-2} 
\left( \partial^i \varphi_0 
\partial_i \varphi_0 \right)- 3 \nabla^{-4} \partial_i \partial^j
\left(\partial^i \varphi_0 \partial_j \varphi_0 \right) \right)\Bigg] 
\nonumber \\
&+& {\mathcal H}^{-2}_0 \left( f_0+\frac{3}{2} \Omega_{0m} \right)^{-1} \Bigg[
g(\eta) \int_{\eta_m}^{\eta} d\tilde{\eta} \, a(\tilde{\eta}) 
{\mathcal H}(\tilde{\eta}) 
S(\tilde{\eta}) - \frac{{\mathcal H}(\eta)}{a^2(\eta)} \int_{\eta_m}^{\eta} 
d\tilde{\eta}\,  a^3(\tilde{\eta}) g(\tilde{\eta}) S(\tilde{\eta})  
\Bigg] \, , \\
\Phi(\eta) & = & \Psi(\eta)+4 g^2(\eta) \varphi_0^2+\frac{4}{3} g^2(\eta) 
\left(\frac{f^2}{\Omega_m}+\frac{3}{2} \right) \left[ 
\nabla^{-2} 
\left( \partial^i \varphi_0 
\partial_i \varphi_0 \right)- 3 \nabla^{-4} \partial_i \partial^j
\left(\partial^i \varphi_0 \partial_j \varphi_0 \right)  \right] \, .
\end{eqnarray} 
Notice that we can rewrite the solutions for $\Psi$ and $\Phi$ in a 
compact form as
\begin{eqnarray}
\label{PSI}
\Psi(\eta)&=&\left( 
B_1(\eta)-2g(\eta)g_{m} -\frac{10}{3}(a_{\rm nl}-1)g(\eta)g_{m}
\right)\varphi_0^2 
+\left( B_2(\eta) -\frac{4}{3}g(\eta)g_{m}  \right) \Bigg[ \nabla^{-2} 
\left( \partial^i \varphi_0 
\partial_i \varphi_0 \right)- 3 \nabla^{-4} \partial_i \partial^j
\left(\partial^i \varphi_0 \partial_j \varphi_0 \right) \Bigg]\nonumber \\
&+& B_3(\eta) \nabla^{-2} \partial_i\partial^j(\partial^i \varphi_0 \partial_j 
\varphi_0 )+B_4(\eta) \partial^i \varphi_0 \partial _i\varphi_0 \, ,\\
\label{PHI}
\Phi(\eta)&=&\left( B_1(\eta)+4g^2(\eta)
-2g(\eta)g_{m} -\frac{10}{3}(a_{\rm nl}-1)g(\eta)g_{m}
\right)\varphi_0^2 
+\Bigg[ B_2(\eta)+\frac{4}{3} g^2(\eta) \left( e(\eta)+\frac{3}{2} \right)
-\frac{4}{3}g(\eta)g_{m} \Bigg] \nonumber \\
& \times & \Bigg[ \nabla^{-2} 
\left( \partial^i \varphi_0 
\partial_i \varphi_0 \right) - 3 \nabla^{-4} \partial_i \partial^j
\left(\partial^i \varphi_0 \partial_j \varphi_0 \right) \Bigg] 
+B_3(\eta) \nabla^{-2} \partial_i\partial^j(\partial^i \varphi_0 \partial_j 
\varphi_0 )+B_4(\eta) \partial^i \varphi_0 \partial _i\varphi_0\, ,
\end{eqnarray}
where we have introduced 
$B_i(\eta)={\mathcal H}_0^{-2} \left(f_0+3 \Omega_{0m}/2 \right)^{-1} 
\tilde{B}_i(\eta)$ with the following definitions
\begin{eqnarray}
\label{B1B2}
\tilde{B}_1(\eta)&=&\int_{\eta_m}^\eta d\tilde{\eta} \,{\mathcal H}^2(\tilde{\eta}) 
(f(\tilde{\eta})-1)^2 C(\eta,\tilde{\eta})\, , \,\,\,\,\,\,\,
\tilde{B}_2(\eta)=2\int_{\eta_m}^\eta d\tilde{\eta} \, {\mathcal H}^2(\tilde{\eta}) 
\Big[2 (f(\tilde{\eta})-1)^2-3
+3 \Omega_m(\tilde{\eta}) \Big] C(\eta,\tilde{\eta})\, , \\
\tilde{B}_3(\eta)&=&\frac{4}{3} \int_{\eta_m}^\eta d\tilde{\eta} \left(e(\tilde{\eta})
+\frac{3}{2} \right) C(\eta,\tilde{\eta}) \, , \,\,\,\,\,\,\,\,\,\,\,\,
\tilde{B}_4(\eta)= - \int_{\eta_m}^\eta d\tilde{\eta} \,C(\eta,\tilde{\eta})\, ,
\end{eqnarray}
and 
\begin{equation}
C(\eta,\tilde{\eta})= g^2(\tilde{\eta}) a(\tilde{\eta}) 
\Big[ g(\eta){\mathcal H}(\tilde{\eta})-g(\tilde{\eta}) 
\frac{a^2(\tilde{\eta})}{a^2(\eta)} {\mathcal H}(\eta) \Big] \, ,
\end{equation}
with $e(\eta)=f^2(\eta)/\Omega_m(\eta)$.

Let us comment on the results found here, in order to proceed further by 
computing the radiation transfer function for the second-order CMB 
anisotropies on large scales. 
In the expression for the second-order gravitational potentials
~(\ref{PSI}) and~(\ref{PHI}) we recognize two contributions. The term which  
dominates on small scales, $[B_3(\eta) \nabla^{-2} 
\partial_i\partial^j(\partial^i \varphi_0 \partial_j 
\varphi_0 )+B_4(\eta) \partial^i \varphi_0 \partial _i\varphi_0]$, which 
gives rise to the second-order Newtonian piece and is insensitive to any
non-linearity in the initial conditions. In fact, for 
a vanishing cosmological constant $\Lambda \rightarrow 0$,  
it is just this term which is responsible for the Rees-Sciama effect 
due to the non-linear evolution of the gravitational potentials. 
For the Einstein-de Sitter case, the signatures of the Newtonian 
Rees-Sciama effect in terms of the CMB bispectrum have been widely 
discussed in the literature~\cite{MGML,munshietal}. 
An analysis of the CMB bispectrum induced by the Newtonian 
Rees-Sciama effect in the presence 
of a non-vanishing cosmological constant $\Lambda \neq 0$ 
has been given in Refs.~\cite{GS,SG}.

The remaining pieces in Eqs.~(\ref{PSI}) and~(\ref{PHI}) 
represent the novelty of our work. They correspond to contributions 
which tend to dominate on large scales with respect to those 
characterizing the 
Newtonian contribution, and whose origin is purely relativistic. 
The contribution of these terms to the CMB anisotropies through 
Eq.~(\ref{main0}) will be mainly on 
large scales through the time dependence of the growth-suppression 
factor $g(\eta)$ and it corresponds to the late ISW effect at second order.   
Moreover these are the pieces carrying the information on primordial 
non-Gaussianity, since they properly take into account 
in the initial conditions the primordial non-Gaussianity generated 
during inflation, as in the standard scenario, or immediately after it, 
as in the curvaton scenario. 
Terms of this type have been obtained for a flat matter-dominated 
(Einstein-de Sitter) Universe  in Refs.~\cite{MMB,mm}, 
but in this case there is no large-scale contribution through the 
integrated effect~({\ref{main0}), 
because the different terms are constant in time, since $g(\eta)=1$ and 
$B_1(\eta)=B_2(\eta)=0$. In Ref.~\cite{tomita1} second-order cosmological 
perturbations have been computed in the 
$\Lambda \neq 0$ case from the synchronous to the Poisson gauge, 
thus extending the analysis of Ref.~\cite{MMB}, and the CMB temperature 
anisotropies induced by metric perturbations have been also considered by 
applying the expressions of Ref.~\cite{mm}.  
However, an important point to notice is that 
both Refs.~\cite{MMB,mm} and Ref.~\cite{tomita1} 
disregard any primordial non-linear contribution 
from inflation.~\footnote{The results 
in Refs.~\cite{MMB,mm,tomita1} have initial conditions corresponding 
to our $a_{\rm nl}=0$.}

\section{Second-order ISW effect in a $\Lambda$CDM Universe: 
primordial non-Gaussianity and second-order radiation transfer function}
\label{SISW}

\subsection{Temperature anisotropies}
\label{TA}

In order to give an expression for the second-order Integrated 
Sachs-Wolfe effect we will take the large-scale contributions to 
Eq.~(\ref{main0}). Thus in Eqs.~(\ref{PSI}),(\ref{PHI}) 
for the second-order gravitational potentials we will drop the Newtonian 
terms that have already been examined in Refs.~\cite{GS,SG} in connection 
to the Rees-Sciama effect. Accordingly we will take the second-order 
corrections $({\rm first-order})^2$ -- quadratic in the linear 
perturbations -- appearing in Eq.~(\ref{main0}) from Refs.~\cite{pyne,mm} 
following the criteria that they must include integrated contributions 
which survive on large scales. 

Let us discuss this step in some details. From Refs.~\cite{pyne,mm} 
we are left to take the large-scale limit of the following expression 
\begin{eqnarray}
\label{corr}
({\rm first-order})^2\equiv\int_{\eta_*}^{\eta_0}  
d\eta \left[ 4 k^{(1)0} \varphi'+4 \varphi' \varphi +2 x^{(1)0} \varphi'' 
+2 x^{(1)i} \varphi'_{,i}\right]
-I_1(\eta_*)(\varphi_*+\tau_{1*}-I_1(\eta_*))\, .
\end{eqnarray}
In Eq.~(\ref{corr}) $\tau_{1*}$ is the linear fractional intrinsic 
temperature fluctuation on the 
last scattering surface, and $I_1(\eta_*)$ is minus the linear ISW effect, 
with
\begin{equation}
\label{I_1}
I_1(\eta)=2 \int_{\eta_0}^{\eta} d\tilde{\eta} \varphi' \, ,
\end{equation}
while $k^{(1)0}$ and $x^{(1)\mu}$ are the first-order perturbation of 
the photon 
wavevector and background geodesic respectively~\cite{pyne,mm}
\begin{eqnarray}
\label{k10}
k^{(1)0}(\eta)&=&-2\varphi+I_1(\eta)\, , \\
\label{x10}
x^{(1)0}(\eta)&=&2 \int_{\eta_0}^{\eta} d\tilde{\eta}
[-\varphi+(\eta-\tilde{\eta}) \varphi']\, , \\
\label{x1i}
x^{(1)i}(\eta)&=& -2 \int_{\eta_0}^{\eta_*} d \eta\, 
[\varphi e^{i}+(\eta -\tilde{\eta}) \varphi^{,i} ]\, ,
\end{eqnarray}
where $e^{i}=-n^{i}$ is the unit vector specifying the line-of-sight 
direction and we have dropped monopole terms due to the observer. 
In Eqs.~(\ref{corr})-(\ref{x1i}) 
the integrands must be evaluated along the background geodesic 
$(\eta,(\eta_0-\eta) e^i)$, 
{\it e.g.} in Eq.~(\ref{I_1}) $\varphi'\equiv\varphi'(\tilde{\eta},{\bf x}=
-{\hat{\bf n}}(\eta_0-\tilde{\eta}))$, while a $*$ denotes quantities 
evaluated at the last 
scattering surface, {\it e.g.} $\varphi_*\equiv 
\varphi(\eta_*,{\bf x}=-{\hat{\bf n}}(\eta_0-\eta_*))$. 

Next, following Ref.~\cite{tomita1} we notice that Eq.~(\ref{corr}) 
can be further simplified. 
We use the relation 
\begin{equation}
x^{(1)0}+x^{(1)i}e_i=-2 \int_{\eta_0}^\eta d \tilde{\eta} \varphi\, ,
\end{equation}      
and the integration by parts 
\begin{equation}
2 \int_{\eta_0}^{\eta*} d \eta \varphi' I_1(\eta)=\frac{1}{2} 
[I_1(\eta_*)]^2\, .
\end{equation} 
Therefore Eq.~(\ref{corr}) becomes 
\begin{eqnarray}
\label{corr1}
({\rm first-order})^2&=&-4 \int_{\eta_*}^{\eta_0}  d\eta\, 
\left[ \varphi \varphi'+ 
\varphi''  \int_{\eta_0}^\eta d \tilde{\eta} 
\varphi \right] + \frac{1}{2} [I_1(\eta_*)]^2-I_1(\eta_*) 
(\tau_{1*}+\varphi_*) 
-4\varphi'_* \int_{\eta_*}^{\eta_0} 
d \eta (\eta_*-\eta) \varphi\, , 
\end{eqnarray} 
where we used the decomposition of the total derivative along the
background geodesic 
for a generic function $f(\eta, x^i(\eta))$, $f'=\partial f/\partial 
\eta=df/d\eta+\partial_i f e^i$. In the following we will use the 
large-scale solution $\tau_{1*}=-2 \varphi_*/3$.

We are now able to provide an expression for the second-order 
ISW effect from Eq.~(\ref{main0}). It is separated into two parts, the early ISW 
effect due to a non-negligible radiation component at last-scattering, and the 
late ISW due to the expansion growth suprresion in a $\Lambda$ dominated epoch. 
Correspondingly 
we will split each integral into two parts as 
$\int_{\eta_*}^{\eta_0} d \eta=\int_{\eta_*}^{\eta_m} d\eta +
 \int_{\eta_m}^{\eta_0} d\eta$, where $\eta_m$ represents the epoch when full 
matter domination is reached. Before that time the early ISW effect is in 
action. Therefore we write the second-order 
ISW effect from Eq.~(\ref{main0}) as 
\begin{equation}
\frac{\Delta T_2}{T}({\bf n})={\frac{\Delta T_2}{T}}^{\rm early}+
{\frac{\Delta T_2}{T}}^{\rm late}\, ,
\end{equation}
with
\begin{equation}
\label{earlysecond}
\frac{1}{2} {\frac{\Delta T_2}{T}}^{\rm early}= \frac{1}{2} \int_{\eta_*}^{\eta_m} d\eta\, 
 (\Phi+\Psi)'({\bf x},\eta) 
\Big|_{{\bf x}=-\hat{{\bf n}}(\eta_0-\eta)}+({\rm first-order})^2_{\rm early}\, ,
\end{equation}
where the $({\rm first-order})_{\rm early}^2$ corrections are obtained from Eq.~(\ref{corr1}) by keeping 
all those contributions which would vanish in the limit of full matter domination at the 
epoch of last scattering. We will discuss the early ISW effect in details 
in section~\ref{earlysection}. 

In the 
following we will focus on the late ISW effect 
\begin{eqnarray}
\label{ISW2}
\frac{1}{2} {\frac{\Delta T_2}{T}}^{\rm late}&=& 
\frac{1}{2} \int_{\eta_m}^{\eta_0} d\eta\, 
(\Phi+\Psi)'({\bf x},\eta) 
\Big|_{{\bf x}=-\hat{{\bf n}}(\eta_0-\eta)}+({\rm first-order})^2_{\rm late} 
\end{eqnarray}
using the results of Sec.~\ref{lambdasecond}. 
Taking the time derivative of Eqs.~(\ref{PSI}) and~(\ref{PHI}) 
the {\it late 
second-order Integrated Sachs-Wolfe effect} is given by  
\begin{eqnarray}
\label{ISW2nd}
\frac{1}{2} {\frac{\Delta T_2}{T}}^{\rm late}
&=&
\int_{\eta_m}^{\eta_0} d\eta\, 
\left[ 2 \left(-1-\frac{5}{3}(a_{\rm nl}-1)\right)\, g_{m}
 g'(\eta)+ B'_1(\eta)
+ 4 g(\eta) g'(\eta)
\right] \varphi_0^2  
\Bigg|_{{\bf x}=- \hat{\bf n}(\eta_0-\eta)} \nonumber \\
&+& \int_{\eta_m}^{\eta_0} d\eta\, g^2_{m} \overline{B}(\eta)
\Bigg[ \nabla^{-2} 
\left( \partial^i \varphi_0 
\partial_i \varphi_0 \right)- 3 \nabla^{-4} \partial_i \partial^j
\left(\partial^i \varphi_0 \partial_j \varphi_0 \right) \Bigg]
\Bigg|_{{\bf x}=- \hat{\bf n}(\eta_0-\eta)}+({\rm first-order})^2_{\rm late}\, .
\end{eqnarray}
We recall that in 
order to obtain the expression~({\ref{ISW2nd}) 
we have dropped from Eqs.~(\ref{PSI}) and~(\ref{PHI}) 
the terms $[B_3(\eta) \nabla^{-2} 
\partial_i\partial^j(\partial^i \varphi_0 \partial_j 
\varphi_0 )+B_4(\eta) \partial^i \varphi_0 \partial _i\varphi_0]$, which   
give rise to the second-order Newtonian piece and give a
 negligible contribution on large scales, and 
in Eq.~(\ref{ISW2nd}) the $({\rm first-order})_{\rm late}^2$ terms are obtained by taking 
that part 
of Eq.~(\ref{corr1}) which is due to late integrated effects   
\begin{eqnarray}
\label{corrf}
({\rm first-order})^2_{\rm late}&=&-4 \int_{\eta_m}^{\eta_0}  d\eta\, \left[  
\varphi \varphi' +\varphi'' \int_{\eta_0}^\eta d \tilde{\eta} 
\varphi \right]+ \frac{1}{2} [I_1(\eta_m)]^2-\frac{1}{3} 
\varphi_* I_1(\eta_m)\, .
\end{eqnarray}

For simplicity of notation in Eq.~(\ref{ISW2nd}) we have introduced
\begin{equation}
\label{barB}
\overline{B}(\eta)=\left( \frac{B'_2(\eta)}{g^2_{m}}-\frac{4}{3} 
\frac{g'(\eta)}
{g_{m}} \right)+\frac{4}{3} \frac{g'(\eta) g(\eta)}{g^2_{m}}
\left( e(\eta)+\frac{3}{2} \right) +
\frac{2}{3}\frac{g^2(\eta)}{g^2_{m}}\, 
e'(\eta)\, .
\end{equation}
Eq.~(\ref{ISW2nd}) is the starting point to evaluate the CMB 
bispectrum induced by the 
late Integrated Sachs-Wolfe effect which arises from the explicit 
time-dependence of the linear 
gravitational potentials during the late accelerated phase~\cite{fut}. 
A simple but 
important comment about Eq.~(\ref{ISW2nd}) is that the whole 
effect is vanishing in the case of a 
vanishing dark-energy component. Moreover we can identify two fundamental 
contributions. One, proportional to $[-5(a_{\rm nl}-1)/3]$, is  
directly sensitive to the primordial non-Gaussianity set in the 
early Universe from 
inflation. The importance of this term comes from the observation 
that in the case of strong 
non-Gaussianity ($|a_{\rm nl}| \gg 1$) the ISW on large scales can 
be strongly amplified. 
The remaining terms are due to the non-linear evolution of the 
gravitational potentials after inflation and to second order-corrections 
in the temperature 
anisotropies. In the case 
of a high level of primordial non-Gaussianity one would expect these 
terms not to give the dominant 
contribution. However accounting for them is crucial in the computation of the
second-order radiation transfer function for CMB temperature anisotropies, 
as we will discuss later.        
\subsection{Angular decomposition}
\label{AD}
In order to compute the angular CMB bispectrum from the second-order 
ISW
effect let us start from the expression of the (today) observed 
CMB anisotropies expanded into spherical harmonics as it is standard 
in the literature, 
with multipoles $a_{lm}$ given by
\begin{equation}
\label{alm}
a_{\ell m}=\int d^2 \hat{\bf n} \frac{\Delta T(\hat{\bf n})}{T} 
Y^{*}_{\ell m}(\hat{\bf n})\, .
\end{equation}  
In terms of the multipoles $a_{lm}$ the linear and non-linear parts 
of the temperature 
fluctuations correspond to a linear Gaussian part $a^{\rm L}_{lm}$ 
and a non-Gaussian 
contribution $a^{\rm NL}_{lm}$
\begin{equation}
a_{\ell m}=a^{\rm L}_{\ell m}+a^{\rm NL}_{\ell m}\, .
\end{equation}
Before computing $a^{\rm NL}_{lm}$ for the second-order ISW  
effect let us recall 
briefly that at linear order the multipoles $a^{\rm L}_{lm}$ 
are expressed as (see, 
{\it e.g} ~\cite{mabert})
\begin{equation}
\label{almL}
a^{\rm L}_{\ell m}= 4\pi (-i)^l \int \frac{d^3 k}{(2 \pi)^3} 
\phi_{1i} ({\bf k}) 
\Delta^{(1)}_\ell (k) Y^*_{\ell m}(\hat{\bf k})\, ,
\end{equation}  
where $\Delta^{(1)}_\ell (k)$ is the linear radiation transfer 
function which describes the 
relations between the initial fluctuations $\phi_{1i} ({\bf k})$ 
and the observed temperature anisotropies. 
 In the following we take this initial 
value at the epoch of matter dominance $\phi_{1i}({\bf k})
=\varphi_{m}({\bf k})$. For the linear Sachs-Wolfe 
effect acting on large scales (small $k$) 
one finds $\Delta^{(1)}_\ell (k)=j_\ell(k(\eta_0-\eta_*))/3$ 
where $j_\ell(x)$ are spherical Bessel functions of order $\ell$. For smaller 
scales one can compute $\Delta^{(1)}_\ell (k)$ with numerical codes such as 
CMBFAST ~\cite{CMBFAST}. 
Expressing the 
non-linear part $a^{\rm NL}_{\ell m}$ in 
terms of the initial potential fluctuations 
corresponds to obtain a second-order radiation transfer function.    
The computation of the radiation transfer function at second-order in 
perturbation 
theory is very useful in that it allows, 
for example, to see what are 
the effects of the non-linearities (set on 
super-horizon scales) at the recombination 
epoch on different scales of the CMB anisotropy pattern. 
In this paper we are able to compute 
the second-order radiation transfer function 
corresponding to the large angular scales of CMB anisotropies.
For small scales (big $k$) the computation requires 
to solve the non-linear Boltzmann equations for the 
photon-matter fluid. 
 A comment is in order here.
Notice that the primordial non-Gaussianity contribution proportional to 
$[-5(a_{\rm nl}-1)/3]$ will be transferred 
linearly to the CMB anisotropies, in that it is already a 
second-order contribution. Thus the crucial  
information to compute the second-order transfer function is carried by 
the remaining contributions in Eq.~(\ref{ISW2nd}), which so far 
have not been properly taken into 
account.\footnote{The analyses of the CMB bispectrum performed so 
far, as for example in 
Refs.~\cite{ks,k}, adopt just the linear radiation transfer function, unless 
the bispectrum originated by specific secondary effects, such as
Rees-Sciama or Sunyaev-Zel'dovich effects, is considered.} 

Let us sketch how to obtain $a^{\rm NL}_{\ell m}$ for the late ISW effect~(\ref{ISW2nd}) taking as an example 
just the term proportional to the initial non-Gaussianity
$A(\hat{\bf n})= - \int_{\eta_m}^{\eta_0} d\eta\, 
\frac{10}{3}(a_{\rm nl}-1) \, g_{m} g'(\eta) \varphi_0^2
|_{{\bf x}=- \hat{\bf n}(\eta_0-\eta)}$ in Eq.~(\ref{ISW2nd}).   
Let us Fourier expand  $A(\hat{\bf n})$   
\begin{equation}
A(\hat{\bf n}) = \int 
\frac{d^3 k}{(2 \pi)^3}  \int_{\eta_m}^{\eta_0} d \eta\, {\mathcal A}\, 
g_{m} g'(\eta)\, 
[\varphi^2_0]({\bf k}) e^{- i {\bf k}\cdot {\hat{\bf n}}(\eta_0-\eta)}\, ,
\end{equation}            
where we set ${\mathcal A}=-10(a_{\rm nl}-1)/3$ and 
$[\varphi^2_0]({\bf k})$ denotes the Fourier transform of 
$\varphi^2_0({\bf x})$. 
If we now make use of the Legendre expansion $e^{i {\bf k} \cdot {\bf x}}=
\sum_\ell (2\ell+1) i^\ell\, j_\ell(kx) P_\ell(\hat{\bf k} 
\cdot \hat{\bf x})$ we find 
\begin{equation}
\label{An}
A(\hat{\bf n}) = \int  \frac{d^3 k}{(2 \pi)^3} 
\sum_{\ell=0}^{\infty} (-i)^\ell (2\ell+1) 
\int_{\eta_m}^{\eta_0} d \eta\, {\mathcal A}\, 
g_{m} g'(\eta)\, j_\ell(k(\eta_0-\eta))\, [\varphi^2_0]({\bf k}) 
P_\ell(\hat{\bf k} \cdot \hat{\bf n})\, .
\end{equation}
Thus inserting Eq.~(\ref{An}) into Eq.~(\ref{alm}) we get the corresponding  
$a^{\rm NL}_{\ell m}$  
\begin{equation}
a^{\rm NL}_{\ell m}(A)=4 \pi(-i)^\ell \int \frac{d^3 k}{(2\pi)^3} 
{\mathcal A}\, [\varphi^2_{m}]({\bf k})
\Bigg[ \int_{\eta_m}^{\eta_0} d \eta\,\frac{g'(\eta)}{g_{m}} 
\, j_\ell(k(\eta_0-\eta)) 
\Bigg] Y^*_{\ell m}(\hat{\bf k})\, ,
\end{equation}
where we have used that $\varphi_{m}=g_{m} \varphi_0$, 
the relation $4 \pi \sum_{m=-\ell}^l Y^*_{\ell m}(\hat{\bf k}) 
Y_{\ell m}(\hat{\bf n})=(2\ell+1) P_\ell(\hat{\bf k} \cdot \hat{\bf n})$ 
and the 
orthonormality of the spherical harmonics. 
${\mathcal A}\,[\varphi^2_m]({\bf k})$ is a convolution
\begin{equation}
{\mathcal A}\,[\varphi^2_m]({\bf k}) = \frac{1}{(2 \pi)^3} \int d^3 k_1 
d^3 k_2 
\delta^{(3)}({\bf k}_1+{\bf k}_2-{\bf k}) \left[-\frac{10}{3}(a_{\rm nl}-1) 
\right]\, 
\varphi_{m}({\bf k}_1) \varphi_{m}({\bf k}_2)\, . 
\end{equation}

Following similar steps 
one can compute the contribution to $a^{\rm NL}_{\ell m}$ from the other 
terms in 
Eq.~(\ref{ISW2nd}). The extension of our example to most of them is 
straightforward, while  
for some terms in the ${\rm (first-order)}_{\rm late}^2$ 
contribution~(\ref{corrf}) some care must be taken. 
The details of the computation can be found in Appendix B, and here 
we report the final result, namely     
\begin{eqnarray}
\label{almNL}
a^{\rm NL}_{\ell m}&=&4 \pi (-i)^\ell \int \frac{d^3k}{(2 \pi)^3} \left[
K_{0}({\bf k}) \Delta^{0(2)}_{\ell}(k)+K_1({\bf k}) \Delta^{1(2)}_{\ell}(k)+
K_2({\bf k}) \Delta^{2(2)}_{\ell}(k) \right]
 Y^*_{\ell m}(\hat{\bf k})\nonumber \\
&+& (4 \pi)^2 \sum_{L_1 M_1} \sum_{L_2 M_2} (-i)^{L_1+L_2} 
{\mathcal G}^{m M_1 M_2}_{\ell L_1 L_2} \int 
\frac{d^3k_1}{(2 \pi)^3} \frac{d^3k_2}{(2 \pi)^3} 
\varphi_{m}({\bf k}_1) \varphi_{m}({\bf k}_2)\,\,  
\Delta_{L_1 L_2}(k_1,k_2) \,\,
Y_{L_1 M_1}(\hat{\bf k}_1) Y_{L_2 M_2}(\hat{\bf k}_2)\, , \nonumber \\
\end{eqnarray}
where $K_n({\bf k})$ are convolutions in Fourier space expressed 
in terms of some kernels 
$f_n({\bf k}_1,{\bf k}_2,{\bf k})$ as 
\begin{equation}
\label{def:Kn}
K_n({\bf k})=\frac{1}{(2 \pi)^3} \int d^3 k_1 d^3 k_2 
\delta^{(3)}({\bf k}_1+{\bf k}_2-{\bf k}) f_n({\bf k}_1,{\bf k}_2,{\bf k}) 
\varphi_{m}({\bf k}_1) \varphi_{m}({\bf k}_2)\, , 
\end{equation} 
with 
\begin{eqnarray}
\label{f01}
f_0({\bf k}_1,{\bf k}_2,{\bf k})&=&-\frac{5}{3}(a_{\rm nl}-1)-1\, , 
\,\,\,\,\,\,\, 
f_1({\bf k}_1,{\bf k}_2,{\bf k})=1 \\
\label{f2}
f_2({\bf k}_1,{\bf k}_2,{\bf k})&=& 3 \frac{({\bf k}_1 \cdot {\bf k})
({\bf k}_2 \cdot {\bf k})}{k^4}-\frac{{\bf k}_1 \cdot {\bf k}_2}{k^2}\, ,
\end{eqnarray}
${\bf k}$ given by ${\bf k}={\bf k_1}+{\bf k_2}$, and correspondingly 
\begin{eqnarray}
\label{TR0}
\Delta^{0(2)}_{\ell}(k)&=&2 \int_{\eta_m}^{\eta_0} d \eta 
\frac{g'(\eta)}{g_{m}} j_\ell(k(\eta_0-\eta))\, , \\
\label{TR1}
\Delta^{1(2)}_{\ell}(k)&=&\int_{\eta_m}^{\eta_0} d \eta  
\frac{B'_1(\eta)}{g^2_m} 
j_\ell(k(\eta_0-\eta))\, ,\\ 
\label{TR2}
\Delta^{2(2)}_{\ell}(k)&=&-\int_{\eta_m}^{\eta_0}d \eta \bar{B}(\eta) 
j_\ell(k(\eta_0-\eta))\, . 
\end{eqnarray} 
In Eq.~(\ref{almNL}) ${\mathcal G}^{m M_1 M_2}_{\ell L_1 L_2}= 
\int d^2{\bf n} Y_{\ell m}(\hat{\bf n}) 
Y_{L_1 M_1}(\hat{\bf n}) Y_{L_2 M_2}(\hat{\bf n})$ is the Gaunt integral and~\footnote{
In fact for the last term of Eq.~(\ref{TR}), $j_{L_2}(k_2(\eta_0-\eta_*))/3$, corresponding to 
the last term of Eq.~(\ref{corrf}), we are assuming full matter domination 
at last scattering $\eta_*$, 
so that we used $\varphi_* \approx \varphi_m$. However in the final formulae we display in Sec. 
V we account also for the small differrence between $\varphi_*$ and $\varphi_m$.
} 
\begin{eqnarray}
\label{TR}
\Delta_{L_1 L_2}(k_1,k_2)&=&- 4 \int_{\eta_m}^{\eta_0} d \eta \,\, 
\frac{g''(\eta)}{g_m}\,
j_{L_1}(k_1(\eta_0-\eta))
\int_{\eta_0}^{\eta} d \tilde{\eta} \frac{g(\tilde{\eta})}{g_m}\, 
j_{L_2}(k_2(\eta_0-\tilde{\eta})) \nonumber \\
&+&2 \int_{\eta_m}^{\eta_0} d\eta\, 
\frac{g'(\eta)}{g_m} j_{L_1}(k_1(\eta_0-\eta)) 
\Bigg[ 2  \int_{\eta_m}^{\eta_0} d\eta\, 
\frac{g'(\eta)}{g_m} j_{L_2}(k_2(\eta_0-\eta))+
\frac{1}{3} j_{L_2}(k_2(\eta_0-\eta_*)) 
\Bigg]\, . \nonumber \\
\end{eqnarray}  
Eq.~(\ref{almNL}) is clearly the generalization of the linear 
relation~(\ref{almL}) with 
the functions $\Delta^{n(2)}_l(k)$ and $\Delta_{L_1L_2}(k_1,k_2)$ 
playing the role of coefficients of the 
{\it second-order radiation transfer function} relating quadratic curvature 
perturbations to the observed temperature anisotropies.
\footnote{Notice that it is possible to rewrite Eq.~(\ref{almNL}) with a more 
general formula as
\begin{equation}
a^{\rm NL}_{\ell m}=
(4 \pi)^2 \sum_{L_1 M_1} \sum_{L_2 M_2} (-i)^{L_1+L_2} 
{\mathcal G}^{m M_1 M_2}_{\ell L_1 L_2} \int 
\frac{d^3k_1}{(2 \pi)^3} \frac{d^3k_2}{(2 \pi)^3} 
\varphi_{m}({\bf k}_1) \varphi_{m}({\bf k}_2)\,\,  
\Delta^{\ell}_{L_1 L_2}({\bf k}_1,
{\bf k}_2) \,\,
Y_{L_1 M_1}(\hat{\bf k}_1) Y_{L_2 M_2}(\hat{\bf k}_2)\, ,
\end{equation} 
with
\begin{equation}
\Delta^{\ell}_{L_1 L_2}({\bf k}_1,{\bf k}_2)=\left[  (-1)^{\ell+L_1+L_2}\, 
\frac{2}{\pi} \,\int 
dk k^2 \sum_{n=0}^2 f_n({\bf k}_1, {\bf k}_2, k) \Delta^{n(2)}_\ell(k) 
\int dr r^2 j_{L_1}(k_1 r) j_{L_2}(k_2 r) j_{\ell}(k r)\right] 
+\Delta_{L_1 L_2}(k_1,k_2)\, . 
\end{equation}
}
For the second-order ISW effect the transfer function depends on the growth 
suppression factor $g(\eta)$. As we anticipated, for the initial 
non-Gaussianity, parametrized by $a_{\rm nl}$ and contained in the 
quadratic term 
$K_0({\bf k})$, the transfer function $\Delta^{0(2)}_\ell(k)$ is 
exactly the same as in linear theory, while the 
transfer function becomes non trivial for the remaining terms quadratic 
in the potential 
perturbations. In particular the part of $a^{\rm NL}_{\ell m}$ expressed 
in terms of the function
$\Delta_{L_1L_2}(k_1,k_2)$ comes from the last three terms in 
Eq.~(\ref{corrf}). 
Notice that for these contributions 
the dependence of $\Delta T/T$ on ${\hat{\bf n}}$ does not enter 
only through a single angle 
$\hat{\bf k} \cdot \hat{\bf n}$, as it is usual in linear theory.      

\subsection{Calculation of the early Integrated Sachs-Wolfe 
effect at second order}
\label{earlysection}
At the epoch $\eta_*$ of last scattering the universe is still not completely matter 
dominated, and the residual radiation component makes the potentials decay in time. 
Such an effect gives a non negligible contribution to the large-scale CMB 
anisotropies, since it almost mimics a Sachs-Wolfe effect. For example at the linear level 
(for a vanishing $\Lambda$) the early ISW effect gives a correction to the power spectrum 
normalization of the order of $20 \%$~\cite{husu95}. A simple estimate from 
Eq.~(\ref{corr1}) shows that also at second-order the early integrated effects must be taken 
into account giving a significant correction to the second-order 
Sachs-Wolfe effect~(\ref{SWa}). 

We consider a universe filled by radiation with energy density $\rho_\gamma$ and pressureless matter with 
energy density $\rho_m$, 
supposing that the cosmological constant term is negligible at the epoch of last scattering.
In this case the linear growing mode of the 
gravitational potential is
\begin{equation}
\label{varphimr}
\varphi = \varphi_* \frac{F(\eta)}{F_*}\, ,
\end{equation}
with~\cite{kodamasasaki,ks1,HuEis,dodelsonbook}
\begin{equation}
\label{defF}
F(\eta)=1-\frac{H}{a} \int_{a_i}^a \frac{da}{H} = \frac{3}{5}+\frac{2}{15y}-\frac{8}{15 y^2}
-\frac{16}{15y^3}+\frac{16\sqrt{1+y}}{15y^3}\, ,
\end{equation}   
where
\begin{equation}
y \equiv \frac{\rho_m}{\rho_\gamma} \propto a\, ,
\end{equation}
$H(a)$ is the Hubble expansion rate, $a_i$ is some  
epoch when radiation is dominating over matter and the $*$ indicates evaluation at the time of 
last scattering. Let us briefly recall how to compute the 
linear early Integrated Sachs-Wolfe effect 
\begin{equation}
\label{almLearly}
a^{\rm L}_{\ell m}= 4\pi (-i)^l \int \frac{d^3 k}{(2 \pi)^3} 
\left[ 2 \int_{\eta_*}^{\eta_m} d\eta \varphi'({\bf k},\eta) j_\ell(k(\eta_0-\eta))\right] 
Y^*_{\ell m}(\hat{\bf k})\, ,
\end{equation}
where the time derivative of the gravitational potential is due to the transition from a 
residual radiation component to full matter domination reached at the epoch $\eta_m$. Since 
most of the contribution to the early ISW effect comes from near recombination at $\eta_*$ the 
integral~(\ref{almLearly}) can be computed by evaluating the Bessel function at $\eta_*$ leaving
\cite{husu95}
\begin{equation}
\label{ap}
2 \int_{\eta_*}^{\eta_m} 
d\eta\, \varphi'({\bf k},\eta) \approx 2\, \varphi({\bf k},\eta)|^{\eta_m}_{\eta_*}\,\, 
j_\ell(k(\eta_0-\eta_*))=2\, \varphi_*({\bf k}) \frac{\Delta F}{F_*}\,\,
j_\ell(k(\eta_0-\eta_*)) \, ,
\end{equation}
where in the last line we used Eq.~(\ref{varphimr}) with 
$\Delta F= F(\eta)|^{\eta_m}_{\eta_*}$. 

Let us now consider the effect at second-order. The early ISW 
effect reads
\begin{equation}
\label{earlysecond}
{\frac{\Delta T_2}{T}}^{\rm early}= \frac{1}{2} \int_{\eta_*}^{\eta_m} d\eta\, 
(\Phi+\Psi)'({\bf x},\eta) 
\Big|_{{\bf x}=-\hat{{\bf n}}(\eta_0-\eta)}+({\rm first-order})^2_{\rm early}\, ,
\end{equation}
where the $({\rm first-order})^2_{\rm early}$ corrections are 
obtained from Eq.~(\ref{corr1}) by keeping 
all those contributions which would vanish in the limit of full matter domination at the 
epoch of last scattering
\begin{eqnarray}
\label{first2early}
({\rm first-order})^2_{\rm early}&=&-4 \int_{\eta_*}^{\eta_m}  d\eta\, 
\left[ \varphi \varphi'+ 
\varphi''  \int_{\eta_0}^\eta d \tilde{\eta} 
\varphi \right] + \frac{1}{2} I^2_1(\eta_m;\eta_*)+I_1(\eta_m;\eta_*)I_1(\eta_*) 
-\frac{1}{3} \varphi_* I_1(\eta_m;\eta_*) 
\nonumber \\
&-&4\varphi'_* \int_{\eta_*}^{\eta_0} 
d \eta (\eta_*-\eta) \varphi\, ,
\end{eqnarray}
where we have splitted $I_1(\eta_*)$ defined in Eq.~(\ref{I_1}) as 
\begin{equation}
I_1(\eta_*)=-2 \int_{\eta_*}^{\eta_0} d\eta \varphi'= -2 \int_{\eta_*}^{\eta_m} d\eta \varphi'
-2 \int_{\eta_m}^{\eta_0} d\eta \varphi' \equiv I_1(\eta_m;\eta_*)+I_1(\eta_m)\, . 
\end{equation} 
Notice that, for example, the last term in Eq.~(\ref{first2early}) proportional to $\varphi_*'$ 
would give a vanishing contribution in the limit of matter domination when the gravitational 
potential $\varphi$ is constant in time.   

For the moment let us focus on the first term on the R.H.S of Eq.~(\ref{earlysecond}). 
The corresponding multipoles are easily computed as in Eq.~(\ref{almLearly})
\begin{equation}
\label{almNLearlys}
a_{\ell m}^{\rm NL} =4\pi (-i)^l \int \frac{d^3 k}{(2 \pi)^3} 
\left[ \frac{1}{2} \int_{\eta_*}^{\eta_m} d\eta  (\Phi+\Psi)'({\bf k},\eta) \,\,
j_\ell(k(\eta_0-\eta))\right] 
Y^*_{\ell m}(\hat{\bf k})\, ,
\end{equation}
with the standard approximation 
\begin{equation}
\label{appearlys}
\frac{1}{2} \int_{\eta_*}^{\eta_m} d\eta  (\Phi+\Psi)'({\bf k},\eta) \,\,
j_\ell(k(\eta_0-\eta)) \simeq \frac{1}{2} (\Phi+\Psi)|^{\eta_m}_{\eta_*}\,\, 
j_\ell(k(\eta_0-\eta_*))\, . 
\end{equation}
Therefore what we need is the expression for the second-order gravitational potentials around the 
epoch of last-scattering when both radiation and matter are considered. The details of such a 
computation are reported in Appendix C, where we consider the evolution of the second-order 
gravitational potentials for a universe filled with radiation and matter. 

First of all we find a relation between the gravitational potential $\Phi$ and $\Psi$ 
\begin{equation}
\label{relPsiPhimr}
\Phi=\Psi+4\varphi^2-{\mathcal Q}\, ,
\end{equation}    
where ${\mathcal Q}$ can be written as 
\begin{equation}
\label{exprQ}
{\mathcal Q} = - \frac{3}{10 F_*^2}\, \widetilde{Q}\, {\mathcal K}\, ,
\end{equation}
with ${\mathcal K}$ given by Eq.~(\ref{calK}) and 
\begin{equation}
\label{tildeQ}
\widetilde{Q}=-4 \frac{1+y}{4+3y} \left[ \dot{F}^2 +2 F\dot{F} +\frac{1}{2} \frac{6+5y}{1+y}F^2 
\right]\, .
\end{equation}
In Eq.~(\ref{tildeQ}) $y=\rho_m/\rho_\gamma$ is the ratio of the matter to the radiation 
component, and a dot stands by a logarithmic derivative with respect to the scale factor, 
$^{\displaystyle{\cdot}} \equiv d/d\ln a$. Eq.~(\ref{relPsiPhimr}) is the analogue of the relation found in 
in Ref.~\cite{BMR2} for a matter-dominated Universe and of Eq.~(\ref{rela}) 
(or Eq.~(\ref{PSI-PHI})) for the case of matter plus a cosmological constant. 

The evolution of the gravitational potentials is then derived from the equation for $\Psi$ 
\begin{equation}
\label{eqPSI}
\frac{\sqrt \rho}{a} \left[ \frac{a}{\sqrt \rho} \Psi \right]^{\displaystyle{\cdot}}=\frac{1}{2} 
\frac{\dot{\rho}}{\rho} \zeta_2+ \left[ \dot{\varphi}^2+{\mathcal Q}
-\frac{1}{2}\frac{\dot{\rho}}{\rho}  \Delta \zeta_2\right]\, , 
\end{equation}
where $\Delta \zeta_2$ is given by Eq.~(\ref{deltaz2}) evaluated for a system of radiation 
plus pressureless matter 
\begin{equation}
\label{exprDz2}
\frac{1}{2} \frac{\dot{\rho}}{\rho} \Delta \zeta_2 = 
\frac{1}{F_*^2} 
\left[ 2 R \dot{F}^2- 2 (2- R) F^2 -4 (1-R) F\dot{F} \right] \varphi^2_*\, ,
\end{equation}
and for simplicity we have introduced the function 
\begin{equation}
R(y)=\frac{1+y}{4+3y}\left( 1+\frac{4}{4+3y} \right)\, .
\end{equation}
Eq.~(\ref{eqPSI}) has been obtained by using the previous relation~(\ref{relPsiPhimr}) 
in the (0-0)-Einstein equation on large scales  
combined with the expression for the second-order curvature perturbation~(\ref{defz2}) 
(see Appendix C). 

In the case of adiabatic perturbations, for which $\zeta_2={\rm constant}$, the 
integration of Eq.~(\ref{eqPSI}) yields
\begin{equation}
\Psi=-F(\eta)\, \zeta_2+\frac{\sqrt \rho}{a} \int_{a_i}^{a} \frac{d a}{\sqrt \rho} 
\left[ \dot{\varphi}^2+{\mathcal Q}
-\frac{1}{2}\frac{\dot{\rho}}{\rho}  \Delta \zeta_2\right]+C \frac{\sqrt{\rho}}{a}\, .
\end{equation}
We thus find the usual linear relation plus a source term quadratic in the 
first-order perturbations. The last term is a decaying mode that can be neglected. 

Finally, using Eq.~(\ref{relPsiPhimr}), we obtain 
\begin{eqnarray}
\label{Psi+phi}
\Psi+\Phi&=&-2 F(\eta)\, \zeta_2+\frac{1}{F_*^2} \left[ 4F^2+2 \frac{\sqrt \rho}{a}
\int_{a_i}^{a} \frac{da}{\sqrt \rho} \left( 
(1-2 R) \dot{F}^2+ 2 (2- R) F^2 +4 (1-R) F\dot{F} \right) \right] \varphi_*^2 
 \nonumber \\
&+&\frac{3}{10 F_*^2}\left[ \widetilde{Q}- 
2 \frac{\sqrt \rho}{a} \int_{a_i}^{a} \frac{da}{\sqrt \rho} \widetilde{Q}
\right] {\mathcal K}\, , 
\end{eqnarray}
which is the expression to be used in Eqs.~(\ref{almNLearlys}) and~(\ref{appearlys}) 
when eavluating the difference 
$(\Psi+\Phi)|^{\eta_m}_{\eta_*}$. The conserved quantity $\zeta_2$ is determined by 
the primordial level of non-Gaussianity parametrized as $\zeta_2=2 a_{\rm nl}\zeta_1^2$, with 
$\zeta_1=-\varphi_*/F_*$. 

To compute the multipoles for the additional terms~(\ref{first2early}) one has to follow the same 
procedure described in Appendix~\ref{B}. We will use Eq.~(\ref{varphimr}) for the epoch when 
radiation is not negligible ({\rm i.e} until the time $\eta_m$)  
and we will adopt the growth suppression 
factor $g(\eta)$ as in Eq.~(\ref{relphiphi_0}) from full matter domination onwards, with 
\begin{equation}
g_{m}\, \varphi_0=\varphi_{m}=\frac{F_m}{F_*} \varphi_*\, ,
\end{equation} 
where $F_m$ denotes the value $F(\eta_m)$. Moreover we will use the evaluation~(\ref{ap}) for the 
linear early ISW effect.

Thus we are able to calculate the radiation transfer 
function for the early ISW effect from Eq.~(\ref{Psi+phi}) and Eq.~(\ref{first2early}). We find
\begin{eqnarray}
\label{almNLE}
a^{\rm NL}_{\ell m}&=&4 \pi (-i)^\ell \int \frac{d^3k}{(2 \pi)^3} \left[
K_{0}({\bf k}) \Delta^{0(2)}_{\ell}(k)+K_1({\bf k}) \Delta^{1(2)}_{\ell}(k)+
K_2({\bf k}) \Delta^{2(2)}_{\ell}(k) \right]
 Y^*_{\ell m}(\hat{\bf k})\nonumber \\
&+& (4 \pi)^2 \sum_{L_1 M_1} \sum_{L_2 M_2} (-i)^{L_1+L_2} 
{\mathcal G}^{m M_1 M_2}_{\ell L_1 L_2} \int 
\frac{d^3k_1}{(2 \pi)^3} \frac{d^3k_2}{(2 \pi)^3} 
\varphi_{*}({\bf k}_1) \varphi_{*}({\bf k}_2)\,\,  
\Delta_{L_1 L_2}(k_1,k_2) \,\,
Y_{L_1 M_1}(\hat{\bf k}_1) Y_{L_2 M_2}(\hat{\bf k}_2)\, , \nonumber \\
\end{eqnarray}
where $\varphi_{*}({\bf k})$ is the gravitational potential at the epoch of last scattering, 
the convolutions $K_n({\bf k})$ are given by      
\begin{equation}
\label{def2:Kn}
K_n({\bf k})=\frac{1}{(2 \pi)^3} \int d^3 k_1 d^3 k_2 
\delta^{(3)}({\bf k}_1+{\bf k}_2-{\bf k}) f_n({\bf k}_1,{\bf k}_2,{\bf k}) 
\varphi_{*}({\bf k}_1) \varphi_{*}({\bf k}_2)\, , 
\end{equation} 
with the kernels $f_n({\bf k}_1,{\bf k}_2,{\bf k})$ defined as in Eqs~(\ref{f01}) and~(\ref{f2}), 
and the transfer functions are given by
\begin{eqnarray} 
\label{earlytr0}
\Delta^{0(2)}_{\ell}(k)&=&\frac{6}{5} \frac{\Delta F}{F^2_*}\, j_{\ell}(k(\eta_0-\eta_*))\, , \\
\Delta^{1(2)}_{\ell}(k)&=&\frac{1}{F_*^2} \left[\frac{\sqrt \rho}{a}
\int_{a_i}^{a} \frac{da}{\sqrt \rho} \left( 
(1-2 R) \dot{F}^2+ 2 (2- R) F^2 +4 (1-R) F\dot{F} \right) \right] \Bigg|_{a_*}^{a_m} 
\, j_{\ell}(k(\eta_0-\eta_*)) \nonumber \\
&+& \frac{4}{5} \frac{\Delta F}{F_*^2}\, j_{\ell}(k(\eta_0-\eta_*))\, \\
\Delta^{2(2)}_{\ell}(k)&=&\frac{1}{F_*^2} \left[ \frac{\widetilde{Q}}{2}- 
\frac{\sqrt \rho}{a} \int_{a_i}^{a} \frac{da}{\sqrt \rho} \widetilde{Q}
\right]\Bigg|_{a_*}^{a_m} \, j_{\ell}(k(\eta_0-\eta_*)) \, ,
\end{eqnarray}
and 
\begin{eqnarray}
\label{earlytrmix}
\Delta_{L_1L_2}(k_1,k_2)&=&\left(2 \frac{\Delta F^2}{F^2_*}  
+\frac{2}{3} \frac{\Delta F}{F_*} \right)\, j_{L_1}(k(\eta_0-\eta_*))\,  
j_{L_2}(k(\eta_0-\eta_*)) +4\frac{\Delta F}{F_*}j_{L_2}(k(\eta_0-{\eta_*}))
\int_{\eta_m}^{\eta_0} d {\eta}\, c\, 
\frac{g'({\eta})}{g_m}\, j_{L_1}(k(\eta_0-{\eta})) 
\nonumber \\
&-& \frac{F'(\eta_*)}{F_*}\, j_{L_2}(k(\eta_0-\eta)) \left[ \int_{\eta_*}^{\eta_m} d\eta
(\eta_*-\eta) \frac{F(\eta)}{F_*} j_{L_1}(k(\eta_0-{\eta_*}))+ 
\int_{\eta_m}^{\eta_0} d\eta
(\eta_*-\eta)\, c\, \frac{g(\eta)}{g_m} j_{L_1}(k(\eta_0-{\eta})) \right]
\nonumber \\
&+&4
\int_{\eta_*}^{\eta_m} d\eta 
\frac{F''(\eta)}{F_*} j_{L_2}(k(\eta_0-\eta)) \left[\int_\eta^{\eta_m} d\tilde{\eta} 
\frac{F(\tilde{\eta})}{F_*} j_{L_1}(k(\eta_0-\tilde{\eta}))+\int_{\eta_m}^{\eta_0} d\tilde{\eta}\,
c\, \frac{g(\tilde{\eta})}{g_m} j_{L_1}(k(\eta_0-\tilde{\eta}))\right]  
\end{eqnarray}
where $c \equiv F_m/F_*$ and 
\begin{equation}
\Delta F = F(\eta_m)-F(\eta_*)\, .
\end{equation}
From Eq.~(\ref{earlytr0})--(\ref{earlytrmix}) we see that the second-order early ISW effect can 
be approximated on large scales as a Sachs-Wolfe effect (whose transfer 
function is porportional to $j_{\ell}(k(\eta_0-\eta_*))$).   
\section{CMB anisotropies from second-order Tensor perturbations}
\label{Tensor2}
The second-order 
tensor contribution to the CMB temperature anisotropies is given by the integrated effect~\cite{mm}
\begin{equation}
\label{ISWT}
\frac{\Delta T_2}{T}({\bf n})=-\frac{1}{2} \int_{\eta_*}^{\eta_0} d\eta\, 
{\chi^{(2)}_{ij}}^{\prime}({\bf x},\eta)\, n^i n^j\Big|_{{\bf x}=- \hat{\bf n}(\eta_0-\eta)}\, .
\end{equation}
On large (superhorizon) scales the tensor modes remain constant (see, {\it e.g.}, Ref.~\cite{SB90}),  
while they start to evolve when they are reentering the 
cosmological horizon. Therefore the main effect on the CMB anisotropies comes from late times, 
and therefore we can take $\eta_* \approx \eta_m$. 
Notice also that the second-order tensor modes are determined by a contribution produced during inflation and 
an additional contribution generated by the post-inflationary evolution.  
Since the level produced during inflation is very small (of the order of the slow-roll 
parameters) compared to the 
post-inflationary contribution, as shown for example in Ref.~\cite{maldacena}, 
we will neglect it in the following.

For a nonvanishing cosmological constant $\Lambda$ and pressureless matter we show 
in Appendix~\ref{A} how to obtain the evolution equation for tensor modes
\begin{equation}
\label{tensorequation}
{{\chi}^i_{2j}}^{\prime \prime} + 2{\cal H} 
{{\chi}^i_{2j}}^{\prime} 
- \nabla^2  \chi^i_{2j}
= \widetilde{{\cal T}}^i_{~j}\, ,
\end{equation}
where the source term is given by 
\begin{equation}
\widetilde{{\cal T}}^i_{~j}
= 4 \nabla^{-2} \partial^k 
\partial_\ell R^\ell_{~k} \delta^i_j 
+ 8  \nabla^{-2}\left(\nabla^2 R^i_{~j} 
- \partial^i \partial_k R^k_{~j} - \partial^k \partial_j 
R^i_{~k} \right) + 4  \nabla^{-4} \partial^i \partial_j \partial^k \partial_\ell 
R^\ell_{~k} 
\;, 
\end{equation}
and the traceless tensor  
\begin{eqnarray}
R^\ell_{~k} & \equiv & \partial^\ell \varphi \partial_k \varphi 
- \frac{1}{3} \left( \nabla \varphi \right)^2 \delta^\ell_k 
+ 4 \pi G a^2 \bar \rho \left( v^\ell_{1} v_{1k} - 
\frac{1}{3} v_{1}^2 \delta^\ell_k \right) \nonumber \\
& = & g^2 \left(1 + \frac{2E^2(z) f^2(\Omega_m)}{3 \Omega_{0m} 
\left(1+z\right)^3}
\right) \left( \partial^\ell \varphi_0 \partial_k  \varphi_0 
- \frac{1}{3} \left( \nabla \varphi_0 \right)^2 \delta^\ell_k \right)
\;.
\end{eqnarray}
It proves convenient to rewrite the source term as 
\begin{equation}
\widetilde{{\cal T}}^i_{~j}=s(\eta) {\cal T}^i_{~j}({\bf k})\, ,
\end{equation}
where 
\begin{equation}
s(\eta)=-8g^2(\eta)\left( 1+\frac{2E^2(z)f^2(\Omega_m)}{3\Omega_{0m}(1+z)^3}\right)\, ,
\end{equation}
and 
\begin{equation}
\nabla^2 {\cal T}_{ij}={\cal S}_{ij} \equiv \nabla^2 \Theta_0 \delta_{ij} + \partial_i 
\partial_j \Theta_0 + 2\left( \partial_i \partial_j 
\varphi_0 \nabla^2\varphi_0 - \partial_i \partial_k
\varphi_0 \partial^k \partial_j \varphi_0\right)\, ,
\end{equation}
with $\Theta_0$ defined in Eq.~(\ref{eq:Theta}).

Let us first consider the solution of the second-order tensor modes.
We can Fourier transform $\chi^i_{2j}$ and the source term 
$\widetilde{\cal T}^i_{~j}$ and decompose them into the $\sigma=+,\times$ polarization modes, 
\begin{eqnarray}
\chi^i_{2j}({\bf x},\eta)&=&\int \frac{d^3{\bf k}}{(2\pi)^3} e^{i{\bf k} \cdot{\bf x}} 
\chi^i_{2j}({\bf k},\eta)\, , \\
\chi^i_{2j}({\bf k},\eta)&=&\chi_+({\bf k},\eta) \epsilon^{+i}_{~~j}(\hat{\bf k})+
\chi_\times({\bf k},\eta) \epsilon^{\times i}_{~~j}(\hat{\bf k})\, ,
\end{eqnarray}
where $\epsilon^{\sigma i}_{~~j}(\hat{\bf k})$  are the polarization tensors, and similarly for the source term.
It is then easy to show that the solution of Eq.~(\ref{tensorequation}) can be written as 
\begin{equation}
\label{solutiontensor}
\chi^i_{2j}({\bf k},\eta)={\cal T}^i_{~j}({\bf k})\, h(k,\eta)\, ,
\end{equation}
where 
\begin{equation}
h(k,\eta)= 
\chi_1(k,\eta) \int_{\eta_m}^{\eta} d\tilde{\eta} \frac{\chi_2(k,\tilde{\eta})}{W} s(\tilde{\eta})
-\chi_2(k,\eta) \int_{\eta_m}^{\eta} d \tilde{\eta} \frac{\chi_2(k,\tilde{\eta})}{W} s(\tilde{\eta})\, , 
\end{equation}
and $\chi_1$ and $\chi_2$ are the solutions for the linear gravity-waves amplitude
\begin{equation}
\label{eveq}
{\chi}^{\prime \prime} + 2{\cal H} {\chi}^{\prime} +k^2 \chi=0 \, ,
\end{equation}
$W=\chi_1'\chi_2-\chi_1\chi_2'$ being their corresponding Wronskian. In Eq.~(\ref{solutiontensor}) 
${\cal T}^i_{~j}({\bf k})$ represents the amplitude of the second-order tensor modes fixed by the source term 
(a quadratic combination of linear scalar modes). 

Let us now consider the expression for the multipole coefficients $a_{\ell m}$ for tensor perturbations. 
We follow some standard References~\cite{Kosowsky,White,KKS} where 
the reader can find more details on the calculations. 
From Eq.~(\ref{alm}) we can define for a given ${\bf k}$-mode the coeffiecients
\begin{equation}
a_{\ell m}({\bf k})=\int d^2{\bf n} \frac{\Delta T({\bf k},\hat{\bf n})}{T} Y^*_{\ell m}(\hat{\bf n})\, ,
\end{equation}
where we have expanded in Fourier space the anisotropy 
$\Delta T({\bf x},\hat{\bf n},\eta)/T= (2\pi^3)^{-1} \int d^3{\bf k}\, e^{i{\bf k}\cdot {\bf x}} \
\Delta T({\bf k},\hat{\bf n},\eta)/T$ to be evaluated today and at the origin. Thus from Eq.~(\ref{ISWT}) we get 
for the $+$ polarization state,
\begin{equation}
\label{alm+}
a^+_{\ell m}({\bf k})=(-1)^{\ell} \int d^2{\hat{\bf n}}\, Y^*_{\ell m}({\hat{\bf n}}) \,
\left[ \frac{1}{4} \int_{\eta_0}^{\eta_m} d\eta
\chi'_+({\bf k},\eta) \right]  \epsilon^+_{ij}(\hat{\bf k})n^i n^j \sum_{\ell'}i^{\ell'} (2\ell'+1)
j_{\ell'}(k(\eta_0-\eta))P_{\ell'}(\hat{\bf n})\, ,
\end{equation} 
where we have used the Legendre expansion $e^{i {\bf k} \cdot {\bf x}}=
\sum_\ell (2\ell+1) i^\ell\, j_\ell(kx) P_\ell(\hat{\bf k} 
\cdot \hat{\bf x})$, and a similar expression holds for the $\times$ polarization state. 

It now proves convenient 
to perform the angular integral for a specific ${\bf k}$-mode, by choosing the coordinate system such that 
$\hat{\bf z}=\hat{\bf k}$. Statistical isotropy of observable quantities, like the angular power spectrum or  the 
angle-averaged bispectrum (and higher-order correlation functions) ensures then that 
we can take the sum 
over the different modes in ${\bf k}$ space at the end. 
Alternatively we will also display a general formula for a generic 
${\bf k}$ (and for the coefficients $a_{\ell m}$). With such a coordinate system choice the integral in 
Eq.~(\ref{alm+}) can be evaluated exactly as shown in Refs.~\cite{White,KKS,Straumann} to give
\begin{eqnarray}
a^+_{\ell m}({\bf k})&=&(-i)^{\ell} \frac{\sqrt \pi}{4} \sqrt{(2\ell+1)} \sqrt{\frac{(\ell+2)!}{(l-2)!}} 
(\delta_{m2}+\delta_{m,-2} )\, 
\int_{\eta_m}^{\eta_0} d\eta \chi'_+({\bf k},\eta) \nonumber \\
&\times& \left[ \frac{j_{\ell+2}}{(2\ell+3)(2\ell+1)}+
2 \frac{j_{\ell}}{(2\ell+3)(2\ell-1)}+\frac{j_{\ell-2}}{(2\ell+1)(2\ell-1)} \right] \nonumber \\
&=& (-i)^{\ell} \frac{\sqrt \pi}{4} \sqrt{(2\ell+1)} \sqrt{\frac{(\ell+2)!}{(l-2)!}}\, 
(\delta_{m2}+\delta_{m,-2} )\, {\cal T}_+({\bf k}) \int_{\eta_m}^{\eta_0} d\eta\, h'(k,\eta) 
\frac{j_{\ell}(k(\eta_0-\eta))}{[k(\eta_0-\eta)]^2}\, ,
\end{eqnarray}
where the spherical Bessel functions are evaluted in $k(\eta_0-\eta)$. In 
the last line we have used Eq.~(\ref{solutiontensor}), ${\cal T}_+({\bf k})$ being the $+$ component of 
the amplitude ${\cal T}^i_{~j}({\bf k})$, and we have applied 
the recursion relation $j_\ell(x)/x=(j_{\ell-1}(x)+j_{\ell+1}(x))/2\ell+1$. 
For the $\times$ polarization state one has to replace $(\delta_{m2}+\delta_{m,-2} )$ with 
$i(\delta_{m,-2}-\delta_{m2})$. 

Following Ref.~\cite{Straumann}, one can also give a formula holding for a generic ${\bf k}$-mode, by using the 
Wigner-D functions~\cite{Vars} to perform a rotation of the spherical harmonics. One thus finds 
\begin{eqnarray}
a_{\ell m}&=& (-i)^{\ell} \frac{\sqrt \pi}{4} \sqrt{(2\ell+1)} \sqrt{\frac{(\ell+2)!}{(l-2)!}}\, 
\int d^3{\bf k} \int_{\eta_m}^{\eta_0} d\eta \frac{j_{\ell}(k(\eta_0-\eta))}{[k(\eta_0-\eta)]^2}
\Big[ (\chi'_+({\bf k},\eta)-i\chi'_{\times} ({\bf k},\eta))   D^{l}_{m2}(\hat{\bf k}) \nonumber \\  
&+& (\chi'_+({\bf k},\eta)+i\chi'_{\times} ({\bf k},\eta))   D^{l}_{m,-2}(\hat{\bf k})  \Big]\, ,
\end{eqnarray}
where $D^l_{mm'}(\hat{\bf k})$ are 
the Wigner-D functions corresponding to a rotation $R(\hat{\bf k})$ which brings 
$\hat{\bf k}$ from $(0,0,1)$ to a generic direction.

The evolution of the second-order tensor perturbations is given by $h(k,\eta)$ having applied the Green method. 
The solutions for the corresponding homogenous equation~(\ref{eveq}) (which is the one for linear tensor modes) 
are well known for the limiting cases $\Omega_m \rightarrow 1$ and $\Omega_\Lambda\rightarrow 1$, and 
correspondingly analytical expression exist for $\chi^i_{2j}$ (see, {\it e.g.},
Ref.~\cite{mhm} and~\cite{CM}). However the full solution requires a numerical evalutation of Eq.~(\ref{eveq}).

\section{Second-order radiation transfer function for large-scale 
CMB anisotropies: Summary table of results and conclusions}

In this section we finalize our computations by providing  
a summary of the formulae for the full radiation transfer function for the 
CMB anisotropies on large angular scales due to the second-order 
Sachs-Wolfe effect in Eq.~(\ref{SWa}), the (early and late) second-order ISW contribution~(\ref{ISW2nd}) and the 
second-order tensor modes, Eq.~({\ref{ISWT}). 
The section can be regarded as a summary of the main results of this paper.  

\subsection{Large-scale CMB anisotropies from second-order scalar perturbations}
The second-order large-scale CMB anisotropies can be written as  
\begin{eqnarray}
\label{almNLfinal}
a^{\rm NL}_{\ell m}&=&4 \pi (-i)^\ell \int \frac{d^3k}{(2 \pi)^3} \left[
K_{0}({\bf k}) \Delta^{0(2)}_{\ell}(k)+K_1({\bf k}) \Delta^{1(2)}_{\ell}(k)+
K_2({\bf k}) \Delta^{2(2)}_{\ell}(k) \right] Y^*_{\ell m}(\hat{\bf k})\nonumber \\
&+& (4 \pi)^2 \sum_{L_1 M_1} \sum_{L_2 M_2} (-i)^{L_1+L_2} 
{\mathcal G}^{m M_1 M_2}_{\ell L_1 L_2} \int 
\frac{d^3k_1}{(2 \pi)^3} \frac{d^3k_2}{(2 \pi)^3} 
\varphi_{*}({\bf k}_1) \varphi_{*}({\bf k}_2)\,\,  
\Delta_{L_1 L_2}(k_1,k_2) \,\,
Y_{L_1 M_1}(\hat{\bf k}_1) Y_{L_2 M_2}(\hat{\bf k}_2)\, , \nonumber \\
\end{eqnarray}
where $\varphi_{*}({\bf k})$ is the gravitational potential at the epoch of last scattering, 
$K_n({\bf k})$ are convolutions in Fourier space expressed in terms of some kernels 
$f({\bf k}_1,{\bf k}_2,{\bf k})$    
\begin{equation}
\label{def:Knfinal}
K_n({\bf k})=\frac{1}{(2 \pi)^3} \int d^3 k_1 d^3 k_2 
\delta^{(3)}({\bf k}_1+{\bf k}_2-{\bf k}) f_n({\bf k}_1,{\bf k}_2,{\bf k}) 
\varphi_{*}({\bf k}_1) \varphi_{*}({\bf k}_2)\, , 
\end{equation} 
with 
\begin{eqnarray}
\label{f1final}
f_0({\bf k}_1,{\bf k}_2,{\bf k})&=&-\frac{5}{3}(a_{\rm nl}-1)-1\, , \\
\label{f2final}
f_1({\bf k}_1,{\bf k}_2,{\bf k})&=&1 \, ,\\
\label{f2final}
f_2({\bf k}_1,{\bf k}_2,{\bf k})&=& 3 \frac{({\bf k}_1 \cdot {\bf k})
({\bf k}_2 \cdot {\bf k})}{k^4}-\frac{{\bf k}_1 \cdot {\bf k}_2}{k^2}\, ,
\end{eqnarray}
${\bf k}$ given by ${\bf k}={\bf k_1}+{\bf k_2}$. 

In Eq.~(\ref{f1final}) $a_{\rm nl}$ is a 
parameter which specifies the level of primordial non-Gaussianity generated in the early 
Universe 
from inflation (or some other alternative mechanisms for the generation of the cosmological 
perturbations)~\cite{review}. The functions $\Delta^{n(2)}_{\ell}(k)$ and 
$\Delta_{L_1L_2}(k_1,k_2)$ represent the second-order radiation transfer functions, and they are 
given by the sum of the transfer functions for the Sachs-Wolfe effect~(\ref{SWa}) and 
the Integrated Sachs-Wolfe effect~(\ref{ISW2nd}).
\\
\\
\begin{itemize}
\item
{\bf Second-order Sachs-Wolfe effect}
\\
\\
Let us first consider the Sachs-Wolfe effect. Notice that the Fourier 
transform of $3 {\cal K}/10$ corresponds to the convolution $K_2({\bf k})$ defined by 
Eq.~(\ref{def:Knfinal}) and 
Eq.~(\ref{f2final}). Thus it is straightforward to compute the corresponding
 multipoles for the Sachs-Wolfe 
effect by using the same steps described in Section~\ref{SISW}. From 
Eq.~(\ref{alm}) we find
\begin{eqnarray}
\label{TRSW}
\Delta^{0(2)}_{\ell}(k)&=&\frac{1}{3}\,j_\ell(k(\eta_0-\eta_*))\, , \\
\label{TR1SW}
\Delta^{1(2)}_{\ell}(k)&=&\frac{7}{18}\, j_\ell(k(\eta_0-\eta_*))\, ,\\ 
\label{TR2SW}
\Delta^{2(2)}_{\ell}(k)&=&- \frac{1}{3} j_\ell(k(\eta_0-\eta_*)) \, , \\
\label{Tr3SW}
\Delta_{L_1L_2}(k_1,k_2)&\equiv &0\, .
\end{eqnarray} 
\\
\\
\item
{\bf Second-order Early ISW effect}
\\
\\
In Section~\ref{earlysection} we found 
\begin{eqnarray} 
\label{earlytr0final}
\Delta^{0(2)}_{\ell}(k)&=&\frac{6}{5} \frac{\Delta F}{F^2_*}\, j_{\ell}(k(\eta_0-\eta_*))\, , \\
\Delta^{1(2)}_{\ell}(k)&=&\frac{1}{F_*^2} \left[\frac{\sqrt \rho}{a}
\int_{a_i}^{a} \frac{da}{\sqrt \rho} \left( 
(1-2 R) \dot{F}^2+ 2 (2- R) F^2 +4 (1-R) F\dot{F} \right) \right] \Bigg|_{a_*}^{a_m} 
\, j_{\ell}(k(\eta_0-\eta_*)) \nonumber \\
&+& \frac{4}{5} \frac{\Delta F}{F_*^2}\, j_{\ell}(k(\eta_0-\eta_*))\, \\
\Delta^{2(2)}_{\ell}(k)&=&\frac{1}{F_*^2} \left[ \frac{\widetilde{Q}}{2}- 
\frac{\sqrt \rho}{a} \int_{a_i}^{a} \frac{da}{\sqrt \rho} \widetilde{Q}
\right]\Bigg|_{a_*}^{a_m} \, j_{\ell}(k(\eta_0-\eta_*)) \, ,
\end{eqnarray}
and 
\begin{eqnarray}
\label{earlytrmixfinal}
\Delta_{L_1L_2}(k_1,k_2)&=&\left(2 \frac{\Delta F^2}{F^2_*}  
+\frac{2}{3} \frac{\Delta F}{F_*} \right)\, j_{L_1}(k(\eta_0-\eta_*))\,  
j_{L_2}(k(\eta_0-\eta_*)) \nonumber \\
&+&4
\left[ \int_{\eta_m}^{\eta_0} d {\eta}\, c\, 
\frac{g'({\eta})}{g_m}\, j_{L_1}(k(\eta_0-{\eta})) \right] 
\frac{\Delta F}{F_*}j_{L_2}(k(\eta_0-{\eta_*}))
\nonumber \\
&+&4
\int_{\eta_*}^{\eta_m} d\eta 
\frac{F''(\eta)}{F_*} j_{L_2}(k(\eta_0-\eta)) \left[\int_\eta^{\eta_m} d\tilde{\eta} 
\frac{F(\tilde{\eta})}{F_*} j_{L_1}(k(\eta_0-\tilde{\eta}))+\int_{\eta_m}^{\eta_0} 
d\tilde{\eta}\,
c\, \frac{g(\tilde{\eta})}{g_m} j_{L_1}(k(\eta_0-\tilde{\eta}))\right]
\nonumber \\
&-& \left[ \int_{\eta_*}^{\eta_m} d\eta
(\eta_*-\eta) \frac{F(\eta)}{F_*} j_{L_1}(k(\eta_0-{\eta_*})) \right] 
\frac{F'(\eta_*)}{F_*}\, j_{L_2}(k(\eta_0-\eta_*)) \nonumber \\
&-& \left[ \int_{\eta_m}^{\eta_0} d\eta
(\eta_*-\eta)\, c\, \frac{g(\eta)}{g_m} j_{L_1}(k(\eta_0-{\eta})) \right]
\frac{F'(\eta_*)}{F_*}\, j_{L_2}(k(\eta_0-\eta_*)) \, .  
\end{eqnarray}
\\
\\
\item
{\bf Second-order Late ISW effect}
\\
\\
From the results of Sections~\ref{lambdasecond},~\ref{TA} and~\ref{AD} we have
\begin{eqnarray}
\label{TR0}
\Delta^{0(2)}_{\ell}(k)&=&2 \int_{\eta_m}^{\eta_0} d \eta\, c^2\,
\frac{g'(\eta)}{g_{m}} j_\ell(k(\eta_0-\eta))\, , \\
\label{TR1}
\Delta^{1(2)}_{\ell}(k)&=&\int_{\eta_m}^{\eta_0} d \eta\, c^2\,  
\frac{B'_1(\eta)}{g^2_m} 
j_\ell(k(\eta_0-\eta))\, ,\\ 
\label{TR2}
\Delta^{2(2)}_{\ell}(k)&=&-\int_{\eta_m}^{\eta_0}d \eta \,c^2\, \bar{B}(\eta) 
j_\ell(k(\eta_0-\eta))\, ,
\end{eqnarray} 
and
\begin{eqnarray}
\label{TRfinal}
\Delta_{L_1 L_2}(k_1,k_2)&=&- 4 \int_{\eta_m}^{\eta_0} d \eta \,\, 
c\,\frac{g''(\eta)}{g_m}\,
j_{L_1}(k_1(\eta_0-\eta))
\int_{\eta_0}^{\eta} d \tilde{\eta}\, c\,  \frac{g(\tilde{\eta})}{g_m}\, 
j_{L_2}(k_2(\eta_0-\tilde{\eta})) \nonumber \\
&+&2 \int_{\eta_m}^{\eta_0} d\eta\, c\,
\frac{g'(\eta)}{g_m} j_{L_1}(k_1(\eta_0-\eta)) 
\Bigg[ 2  \int_{\eta_m}^{\eta_0} d\eta\, c\, 
\frac{g'(\eta)}{g_m} j_{L_2}(k_2(\eta_0-\eta))+
\frac{1}{3} j_{L_2}(k_2(\eta_0-\eta_*) 
\Bigg]\, . \nonumber \\
\end{eqnarray}  
\end{itemize}

Let us recall the definition of the different symbols and 
quantities  appearing in the previous formulae.
\begin{itemize}
\item[{\bf -}] 
{\bf Symbols}
\begin{eqnarray}
\eta  & & \textrm{conformal time} \nonumber \\
a & & \textrm{scale factor} \nonumber \\
a_i & & \textrm{epoch of radiation domination} \nonumber \\
\eta_* & & \textrm{epoch of last scattering} \nonumber \\
\eta_m & & \textrm{epoch of full matter domination} \nonumber \\
\eta_0 & & \textrm{present epoch} \nonumber \\
\varphi_* & & \textrm{(linear) gravitational potential at the epoch of last scattering} \\
^{\displaystyle{\cdot}}& \equiv & \frac{d}{d \ln a} \nonumber \\
' &\equiv & \frac{d}{d\eta}   \nonumber 
\end{eqnarray}

\item[{\bf -}] 
{\bf Functions}
\item 
$F(\eta)$ gives the linear growing-mode solution for the gravitational potential 
$\varphi$ for 
adiabatic perturbations of pressurelss matter and radiation fluids~\cite{kodamasasaki,ks1,HuEis,dodelsonbook}
\begin{equation}
\label{defF}
F(\eta) = \frac{3}{5}+\frac{2}{15y}-\frac{8}{15 y^2}
-\frac{16}{15y^3}+\frac{16\sqrt{1+y}}{15y^3}\, ,
\end{equation}   
where
\begin{equation}
y \equiv \frac{\rho_m}{\rho_\gamma} \propto a\, ,
\end{equation}
with 
\begin{equation}
c \equiv F_m/F_* 
\end{equation}
and 
\begin{equation}
\Delta F = F(\eta_m)-F(\eta_*)\, .
\end{equation}

\item $g(\eta)$ is the growth-suppression factor for a flat Universe filled with 
a cosmological 
constant $\Lambda$ and a pressureless fluid whose exact form can be found in 
Refs.~\cite{lahav,Carroll,Eisenstein}. 
In the $\Lambda=0$ case $g=1$. An excellent approximation for $g$ as a 
function of redshift $z$ is given in 
Refs.~\cite{lahav,Carroll} 
\begin{equation}
g \propto \Omega_m\left[\Omega_m^{4/7} - \Omega_\Lambda +
\left(1+ \Omega_m/2\right)\left(1+ \Omega_\Lambda/70\right)\right]^{-1} \, , 
\end{equation}
with $\Omega_m=\Omega_{0m}(1+z)^3/E^2(z)$, 
$\Omega_\Lambda=\Omega_{0\Lambda}/E^2(z)$, 
$E(z) \equiv (1+z) {\mathcal H}(z)/{\mathcal H}_0 = \left[\Omega_{0m}(1+z)^3 + 
\Omega_{0\Lambda}\right]^{1/2}$ and 
$\Omega_{0m}$, $\Omega_{0\Lambda}=1-\Omega_{0m}$, the present-day
density parameters of non-relativistic matter and cosmological constant, 
respectively. We have normalized $g(\eta)$ so that at the present epoch 
$g(z=0)=1$. 
The quantity $g_{m}$ corresponds to the value of $g(\eta)$ at the epoch 
$\eta_m$ when full matter domination starts. A good approximation is
\begin{equation}
g_m \approx \frac{2}{5}\, \Omega_{0m}^{-1}\,(\Omega_{0m}^{4/7} + 
\frac{3}{2}\Omega_{0m})\, .
\end{equation}
\item
The function $\widetilde{Q}$ enters in the difference between the two second-order 
gravitational potentials $\Psi$ and $\Phi$, Eq.~(\ref{relPsiPhimr}), and it is given by
\begin{equation}
\label{tildeQfinal}
\widetilde{Q}=-4 \frac{1+y}{4+3y} \left[ \dot{F}^2 +2 F\dot{F} +\frac{1}{2} \frac{6+5y}{1+y}F^2 
\right]\, .
\end{equation}
\item 
The function $R(y)$ has been introduced for simplicty and it is defined as 
\begin{equation}
R(y)=\frac{1+y}{4+3y}\left( 1+\frac{4}{4+3y} \right)\, .
\end{equation}
\item
The functions $B_1(\eta)$ and $B_2(\eta)$ enter in the large-scale solution of the gravitational 
potentials $\Psi$ and $\Phi$, Eqs.~(\ref{PSI}) and~(\ref{PHI}), 
\begin{eqnarray}
\label{B1final}
B_1(\eta)&=&
{\mathcal H}_0^{-2} \left(f_0+3 \Omega_{0m}/2 \right)^{-1} 
\int_{\eta_m}^\eta d\tilde{\eta} \,{\mathcal H}^2(\tilde{\eta}) 
(f(\tilde{\eta})-1)^2 C(\eta,\tilde{\eta})\, , \\
\label{B2final}
B_2(\eta)&=&2
{\mathcal H}_0^{-2} \left(f_0+3 \Omega_{0m}/2 \right)^{-1} 
\int_{\eta_m}^\eta d\tilde{\eta} \, {\mathcal H}^2(\tilde{\eta}) 
\Big[2 (f(\tilde{\eta})-1)^2-3
+3 \Omega_m(\tilde{\eta}) \Big] C(\eta,\tilde{\eta})\, ,
\end{eqnarray}
where
\begin{eqnarray}
C(\eta,\tilde{\eta})&=& g^2(\tilde{\eta}) a(\tilde{\eta}) 
\Big[ g(\eta){\mathcal H}(\tilde{\eta})-g(\tilde{\eta}) 
\frac{a^2(\tilde{\eta})}{a^2(\eta)} {\mathcal H}(\eta) \Big] \, , \\
f(\eta)&=&1+\frac{g'(\eta)}{{\mathcal H}g(\eta)} \, ,\\
e(\eta)&=&\frac{f^2(\eta)}{\Omega_m(\eta)} \, .
\end{eqnarray}
\item ${\bar B}(\eta)$ is defined by
\begin{equation}
\label{barBfinal}
\overline{B}(\eta)=\left( \frac{B'_2(\eta)}{g^2_{m}}-\frac{4}{3} 
\frac{g'(\eta)}
{g_{m}} \right)+\frac{4}{3} \frac{g'(\eta) g(\eta)}{g^2_{m}}
\left( e(\eta)+\frac{3}{2} \right) +
\frac{2}{3}\frac{g^2(\eta)}{g^2_{m}}\, 
e'(\eta)\, .
\end{equation}
\end{itemize}

\subsection{CMB anisotropies from second-order tensor modes}
In Section~\ref{Tensor2} we found that the contribution of a single ${\bf k}$-perturbation mode, 
with the coordinate system choosen with $\hat{\bf z}=\hat{\bf k}$, is given by
\begin{eqnarray} 
a^+_{\ell m}({\bf k})&=& (-i)^{\ell} \frac{\sqrt \pi}{4} \sqrt{(2\ell+1)} \sqrt{\frac{(\ell+2)!}{(l-2)!}}\, 
(\delta_{m2}+\delta_{m,-2} )\, {\cal T}_+({\bf k}) \int_{\eta_m}^{\eta_0} d\eta\, h'(k,\eta) 
\frac{j_{\ell}(k(\eta_0-\eta))}{[k(\eta_0-\eta)]^2}\, , \\
a^\times_{\ell m}({\bf k})&=& (-i)^{\ell} \frac{\sqrt \pi}{4} \sqrt{(2\ell+1)} \sqrt{\frac{(\ell+2)!}{(l-2)!}}\, 
i (\delta_{m,-2}-\delta_{m2} )\, {\cal T}_\times({\bf k}) \int_{\eta_m}^{\eta_0} d\eta\, h'(k,\eta) 
\frac{j_{\ell}(k(\eta_0-\eta))}{[k(\eta_0-\eta)]^2}\, ,
\end{eqnarray} 
for the $+$ and $\times$ polarization states respectively. 

\begin{itemize}
\item
${\cal T}_+({\bf k})$ and  ${\cal T}_\times({\bf k})$ are the $+$ and $\times$ component of the amplitude of 
the second-order tensor modes, 
${\cal T}^i_{~j}({\bf k})= \int d^3{\bf y}\, e ^{-i{\bf y}\cdot{\bf k}} {\cal T}^i_{~j}({\bf y})$ where
\begin{equation}
\nabla^2 {\cal T}_{ij}=\nabla^2 \Theta_0 \delta_{ij} + \partial_i 
\partial_j \Theta_0 + 2\left( \partial_i \partial_j 
\varphi_0 \nabla^2\varphi_0 - \partial_i \partial_k
\varphi_0 \partial^k \partial_j \varphi_0\right)\, ,
\end{equation}
with
\begin{equation}
\label{eq:Thetafinal}
\nabla^2\Theta_0 = -\frac{1}{2}\left((\nabla^2\varphi_0)^2-
\partial_i \partial_k \varphi_0 \partial^i \partial^k 
\varphi_0\right) \, .
\end{equation}

\item
$h(k,\eta)$ describes the evolution of the second-order tensor perturbations 
 \begin{equation}
h(k,\eta)= \chi_1(k,\eta) \int_{\eta_m}^{\eta} d\tilde{\eta} \frac{\chi_2(k,\tilde{\eta})}{W} s(\tilde{\eta})
-\chi_2(k,\eta) \int_{\eta_m}^{\eta} d \tilde{\eta} \frac{\chi_2(k,\tilde{\eta})}{W} s(\tilde{\eta})\, , 
\end{equation}
where 
\begin{equation}
s(\eta)=-8g^2(\eta)\left( 1+\frac{2E^2(z)f^2(\Omega_m)}{3\Omega_{0m}(1+z)^3}\right)\, ,
\end{equation}
and $\chi_1$ and $\chi_2$ are the solutions for the linear tensor modes
\begin{equation}
\label{eveqfinal}
{\chi}^{\prime \prime} + 2{\cal H} {\chi}^{\prime} +k^2 \chi=0 \, ,
\end{equation}
$W=\chi_1'\chi_2-\chi_1\chi_2'$ being their corresponding Wronskian. 

\end{itemize}

\subsection{Conclusions}
In this paper we have computed the full second-order radiation transfer function for 
Cosmic Microwave Background anisotropies on large angular scales 
in a flat universe filled with matter and cosmological constant $\Lambda$. 
Our calculation includes the second-order generalization of the Sachs-Wolfe effect as well as of (both the early 
and the late)  Integrated Sachs-Wolfe 
effects, 
and is valid for a generic set of initial conditions specifying the level of primordial non-Gaussianity. We also 
accounted for the contribution of second-order tensor modes. Our results deal with large angular scales 
and represent the first step towards the 
computation of the full second-order radiation transfer function for 
CMB anisotropies. 
Its computation on small angular scales involves several   
non-linear effects, such as, for example, gravitational lensing, Shapiro time-delay, 
Rees-Sciama effects, and  
of course, dealing with the non-linear Boltzmann equation for the 
photon-matter fluid. Such computation will be the subject of a 
future publication~\cite{fut}.

\section*{Acknowledgments}
We thank Michele Liguori for useful discussions. 
A.~R. is on leave of absence from INFN, Padova (Italy). S.~M. acknowledges partial financial support by INAF. N.~B.
acknowledges partial financial support by INFN.

\vskip 1cm
\appendix
\setcounter{equation}{0}
\def\theequation{A.\arabic{equation}}
\vskip 0.2cm
\section{Perturbations of a flat $\Lambda$CDM Universe in the Poisson gauge.}
\label{A}
\subsection{Background equations}
Let us consider a flat Friedmann-Robertson-Walker (FRW) universe 
filled with a pressurelss fluid of 
energy momentum tensor $T^\mu_{~\nu}= \rho u^\mu u_\nu$ plus a cosmological 
constant $\Lambda$. 
The Friedmann equations are given by 
\begin{eqnarray}
\label{00back}
{\mathcal H}^2&=&\frac{a^2}{3} (8 \pi G \bar{\rho} +\Lambda)\, , \\
\label{contback}
\bar{\rho}'&=&-3 {\mathcal H} \bar{\rho} \, ,
\end{eqnarray}
where $a(\eta)$ is the scale factor, a bar denotes the background quantities 
and ${\mathcal H}=a'/a$, with a prime standing for 
a derivative with respect to conformal time $\eta$. Another useful relation is 
\begin{equation}
\label{u}
2 {\mathcal H}'=-{\mathcal H}^2+a^2 \Lambda\, .
\end{equation} 
Finally let us recall the definitions of the density parameters 
$\Omega_m=8 \pi G a^2 \bar{\rho}/(3 {\cal H}^2)$ and $\Omega_\Lambda=a^2 
\Lambda/(3 {\cal H})^2$ 
for the non-relativistic matter component and the cosmological constant 
respectively, with $\Omega_m+
\Omega_\Lambda=1$. 
\subsection{Linear perturbations}
We briefly report the results for the linear perturbations~(for a 
detailed analysis, see {\it e.g.} Ref.~\cite{fmb}).  
The perturbed metric tensor is defined by Eq.~(\ref{metric}) and we will 
adopt the Poisson gauge for 
which $\omega_i$ is a pure vector, {\it i.e.} $\partial_i \omega^i=0$, while 
$\chi_{ij}$ is a tensor mode ({\it i.e.} divergence-free and traceless     
$\partial^i \chi_{ij}=0$, $\chi^i_{~i}=0$). The matter component has mass density 
$\rho({\bf x},\eta)=\bar{\rho}(\eta)(1+\delta)$ and four-velocity 
$u^{\mu}=(\delta^\mu_0+v^\mu)/a$, with
$u^\mu u_\mu=-1$. Each perturbation is expanded into a first-order 
(linear) and second-order part, 
{\it e.g.} for the mass density perturbation 
$\delta=\delta_1({\bf x},\eta)+\delta_2({\bf x},\eta)/2$.
The traceless part of the $(i-j)$ Einstein's equations in the Poisson
 gauge imply that 
$\phi_1=\psi_1 \equiv \varphi$, while its trace gives the evolution equation
\begin{equation}
\varphi''+3 {\mathcal H} \varphi'+(2{\mathcal H}'+{\mathcal H}^2) \varphi=0\, ,
\end{equation}
which, with Eq.~(\ref{u}), brings to Eq.~(\ref{ev}).
 Vector modes are not generated from standard 
mechanisms for cosmological perturbation, such as inflation. Tensor
 modes obey an 
evolution equation obtained from the traceless part of the 
($i$-$j$)-component of Einstein equations 
\begin{equation}
\label{gravwaves}
{{\chi}^{i}_{1j}}'' + 2{\cal H} {\chi^{i}_{1j}}'
- \nabla^2  \chi^i_{1j} =0 \, .
\end{equation}
However, for the purposes of the present paper, we can neglect 
the linear tensor modes in the 
following.

The energy and momentum constraints provide the density and
velocity fluctuations in terms of $\varphi$ 
(see, for example, Ref.~\cite{BMR2} and~\cite{mhm,tomita1} 
for the $\Lambda$ case)
\begin{eqnarray} 
\label{lindensvel}
\delta_1& = & \frac{1}{4 \pi G a^2 \bar\rho} 
\left[ \nabla^2 \varphi - 3{\cal H}\left(\varphi' + {\cal H} 
\varphi\right) \right], \\
\label{v1}
v_{1i} & = & - \frac{1}{4 \pi G a^2 \bar\rho} \partial_i 
\left(\varphi' + {\cal H} \varphi\right) \;.  
\end{eqnarray}
\subsection{Second-order perturbations}
\subsubsection{Scalar perturbations}
\label{appsc}
Let us first consider the second-order scalar perturbations of 
the metric in the Poisson gauge 
$\phi_2 \equiv \Phi$ and $\psi_2 \equiv \Psi$. 
We will now show how to obtain the evolution equation~(\ref{Psieq}) 
for the gravitational potential $\Psi$ and the relation~(\ref{rela}) which 
represent our master equations. 

The second-order perturbations of the Einstein tensor $G^\mu_{~\nu}$ 
can be found for any gauge in 
Appendix A of Refs.~\cite{ABMR,review}. The perturbations of the 
energy-momentum tensor up to 
second order in the Poisson gauge have been computed in Ref.~\cite{BMR2} 
for a general perfect fluid 
(see Ref.~\cite{review} for expressions in any gauge). For the 
pressureless case one has
\begin{eqnarray}
T^0_{~0} & = & -\bar\rho \left(1+ \delta_1 + \frac{1}{2} \delta_2 + 
v_1^2 \right) \;, \nonumber \\
T^i_{~0} & = &  -\bar\rho \left[ v_1^i + \frac{1}{2} v_{2}^i
+\left(\varphi+\delta_1\right)v_1^i \right]   \;, \nonumber \\
T^0_{~i} & = &  \bar\rho \left[ v_{1i} + \frac{1}{2} v_{2i}
+ \frac{1}{2} \omega_{2i} 
+\left(-3\varphi+\delta_1\right)v_{1i} \right] \;, \nonumber \\
T^i_{~j} & = &  \bar\rho ~ v_{1}^i v_{1j} \, ,
\end{eqnarray}
where $v^2_1=v_1^i v_{1i}$. Note that the second-order 
velocity $v_{2i}$ is the sum of an 
irrotational component $v^{||}_{2i}$, which is the gradient of a 
scalar, and a rotational vector $v^{\bot}_{2i} $, 
which has zero divergence, $\partial^i v^{\bot}_{2i}=0$.  

We thus obtain the ($i$-$j$)-component of Einstein equations
\begin{eqnarray}
\label{ij}
& & \left[ \Psi''+{\cal H} (2 \Psi'+\Phi') +\frac{1}{2} \nabla^2 (\Phi-\Psi)+ 
(2{\cal H}'+{\cal H}^2) \Phi-4 ({\cal H}'+{\cal H}^2)\, 
\varphi^2 -\varphi^{\prime 2} -8 {\cal H}\varphi 
\varphi' -3 \left(\partial_i \varphi \partial^i \varphi \right)  
-4\varphi \nabla^2 \varphi \right] 
\delta^i_j \nonumber \\
& &-  \frac{1}{2} \partial^i \partial_j (\Phi-\Psi)-\frac{1}{2} {\cal H} 
(\partial^i \omega_{2j}+\partial_j \omega^i_2)-\frac{1}{4}(\partial^i 
{\omega_{2j}}^{\prime}+
\partial_j {\omega^{i}_2}^{\prime})+
\frac{1}{4}\left[{\chi^{i}_{2j}}^{\prime \prime} +2 {\cal H} 
{\chi^{i}_{2j}}^{\prime} -\nabla^2 \chi^i_{2j}\right]+2 \partial^i 
\varphi \partial_j \varphi+4 \varphi 
\partial^i \partial_j \varphi = \nonumber \\
& & 8 \pi G a^2 \bar{\rho}\, v^i_1 v_{1j}\, .
\end{eqnarray}
We now proceed as follows. We take the traceless part of Eq.~(\ref{ij}), 
$\delta_2 G^i_{~j} - \frac{1}{3}\delta_2 G^k_{~k} 
\delta^i_j = 8 \pi G (\delta_2 T^i_{~j} - 
\frac{1}{3} \delta_2 T^k_{~k} \delta^i_j)$, namely 
\begin{eqnarray}
\label{tracelessij}
&& - \left[\frac{1}{6}\nabla^2\left(\Psi - \Phi \right) + 
\frac{2}{3} \left(\nabla \varphi\right)^2 + 
\frac{4}{3} \varphi \nabla^2 \varphi \right] 
\delta^i_j + \frac{1}{2} \partial^i \partial_j 
\left(\Psi - \Phi \right) 
+ 2 \partial^i \varphi \partial_j \varphi + 4 \varphi \partial^i \partial_j 
\varphi \nonumber \\
&& 
- \frac{1}{4} \left(\partial^i {{\omega}_{2j}}{\prime}  + \partial_j 
{{\omega}_2^i}^{\prime} \right) - \frac{1}{2} {\cal H} 
\left( \partial^i \omega_{2j}+ \partial_j 
\omega_2^i \right) +
\frac{1}{4} \left( {{\chi}^i_{2j}}^{\prime \prime} + 2{\cal H} 
{{\chi}^i_{2j}}^{\prime} 
- \nabla^2  \chi^i_{2j}\right)  
= 8 \pi G a^2 
\bar\rho \left( v_1^i v_{1j} - \frac{1}{3} v_1^2 
\delta^i_j \right),
\end{eqnarray}
and we apply the operator $\partial_i \partial^j$ in order to get rid 
of second-order vector and tensor 
perturbation modes and to solve for the combination $(\Psi-\Phi)$. We find
\begin{equation}
\label{PSI-PHI}
\Psi-\Phi=-4\varphi^2+{\cal Q}\, ,
\end{equation}
where ${\cal Q}$ is defined by 
\begin{equation}
\label{defQ}
\nabla^2{\cal Q} = -P+3 N\, ,
\end{equation}
with 
\begin{equation}
\label{defP}
P \equiv P^i_{~i}\, ,
\end{equation}
where
\begin{equation}
\label{defPij}
P^i_{~j} = 2 \partial^i \varphi \partial_j \varphi + 
8\pi G a^2 \bar{\rho}\, v^i_1 v_{1j}\, , 
\end{equation}
and the quantity $N$ is given by 
\begin{equation}
\label{defN}
\nabla^2 N \equiv \partial_i \partial^j P^i_{~j}\ .
\end{equation}
The combination $(\Psi-\Phi)$ will be used in the trace of the $(i-j)$ 
Einstein equation that 
we write in the form
\begin{eqnarray}
\label{traceij}
\Psi''+3{\cal H} \Psi'+a^2 \Lambda \Psi &=& {\cal H}(\Psi'-\Phi') 
+a^2 \Lambda (\Psi-\Phi) +
\frac{1}{3} \nabla^2 (\Psi-\Phi) +4 a^2 \Lambda \varphi^2 +\varphi^{\prime 2}
+\frac{1}{3} (\partial^i \varphi 
\partial_i \varphi)+8 {\cal H} \varphi \varphi'+\frac{8}{3} 
\varphi \nabla^2 \varphi \nonumber \\ 
&+&\frac{8 \pi G}{3} \bar{\rho}\, a^2 v^2_{1}\, ,  
\end{eqnarray}
where we have used Eq.~(\ref{u}). Inserting Eq.~(\ref{PSI-PHI}) 
into Eq.~(\ref{traceij}) and 
using Eq.~(\ref{defQ}) and~(\ref{defP}) we find
\begin{eqnarray}
\label{traceijQN}
\Psi''+3{\cal H} \Psi'+a^2 \Lambda \Psi &=&{\cal H}{\cal Q}'+a^2 
\Lambda {\cal Q}+N+\varphi^{\prime 2}
-(\partial^i \varphi \partial_i \varphi)\, .
\end{eqnarray}

The next step is to write explicitly ${\cal H}{\cal Q}'+a^2 
\Lambda {\cal Q}$ and $N$ through 
Eqs.~(\ref{defQ})-(\ref{defN}). Notice that by using Eq.~(\ref{v1})
 one can write $P^i_{~j}$ as 
\begin{eqnarray}
P^i_{~j}=\frac{2}{4\pi G a^2 \bar{\rho}} \left[ \partial^i \varphi' 
\partial_j \varphi' 
+{\cal H}(\partial^i \varphi \partial_j \varphi)' 
+\frac{1}{2} (5 {\cal H}^2-a^2 \Lambda) 
\partial^i \varphi \partial_j \varphi \right]\, ,
\end{eqnarray}  
where we have also used the background equation~(\ref{00back}). 
Thus from Eq.~(\ref{defN}) 
\begin{eqnarray}
\nabla^2\nabla^2N=\frac{2}{4\pi G a^2 \bar{\rho}} 
\nabla^2 \partial_i \partial^j
\left[ \partial^i \varphi' \partial_j \varphi' 
+{\cal H}(\partial^i \varphi \partial_j \varphi)' 
+\frac{1}{2} (5 {\cal H}^2-a^2 \Lambda) 
\partial^i \varphi \partial_j \varphi \right]\, .
\end{eqnarray}  
For the combination ${\cal H}{\cal Q}'+a^2 \Lambda {\cal Q}$, 
with the definitions~(\ref{defQ}) 
and~(\ref{defN}), we can write
\begin{eqnarray}
\nabla^2 \nabla^2 ({\cal H}{\cal Q}'+a^2 \Lambda {\cal Q})=-\nabla^2 
({\cal H} P'+a^2\Lambda P)+
3\partial_i \partial^j ({\cal H}{P^{i}_{~j}}^{\prime} +a^2 
\Lambda P^{i}_{~j})\, .
\end{eqnarray} 
Using the background equations~(\ref{00back}),~(\ref{contback}) 
and the evolution equation~(\ref{ev}) 
for the linear potential $\varphi$ we get
\begin{eqnarray}
{P^i_{~j}}^{\prime}=-\frac{2}{4\pi G a^2\bar{\rho}}\left[ 3 {\cal H} 
\partial^i \varphi' \partial_j \varphi'+a^2 \Lambda {\cal H} \, 
\partial^i \varphi \partial_j \varphi
+a^2\Lambda(\partial^i\varphi'\partial_j\varphi
+\partial^i\varphi\partial_j\varphi')\right]\, , 
\end{eqnarray}
and $ {\cal H}{P^i_{~j}}^{\prime}+a^2\Lambda P^i_{~j}=2 a^2 
\Lambda \partial^i \varphi \partial_j \varphi
-4 \partial^i \varphi' \partial_j \varphi'$. We then find
\begin{eqnarray}
\nabla^2\nabla^2({\cal H}{\cal Q}'+a^2 \Lambda {\cal Q})=4 \left[ 
\nabla^2 (\partial_i \varphi' \partial ^i \varphi')-3 \partial_i \partial^j
(\partial^i \varphi' \partial_j \varphi') \right]  -2a^2 
\Lambda \left[ \nabla^2 
(\partial_i \varphi \partial ^i \varphi) -3 
\partial_i \partial^j
(\partial^i \varphi \partial_j \varphi) \right]\, .
\end{eqnarray}
We can use the relation $f(\Omega_m)\equiv d\ln D_+/d \ln a=
1+ g'(\eta)/({\cal H} g(\eta))$ and Eq.~(\ref{relphiphi_0}) to write
\begin{eqnarray}
\label{Qf}
{\cal H} {\cal Q}'+a^2 \Lambda {\cal Q}= 2 g^2 {\cal H}^2 \Omega_m 
\left(2 \frac{(f(\Omega_m)-1)^2}{\Omega_m} -\frac{3}{\Omega_m} +3 \right) 
\Bigg(\nabla^{-2}(\partial^i \varphi_0 \partial_i \varphi_0)-3 \nabla^{-4} 
\partial^i 
\partial_j(\partial_i \varphi_0 \partial^j \varphi_0) \Bigg)\, ,
\end{eqnarray} 
and 
\begin{eqnarray}
\label{Nf}
N=\frac{4}{3}g^2\left( \frac{f^2(\Omega_m)}{\Omega_m} +\frac{3}{2} \right)
\nabla^{-2} \partial_i \partial^j (\partial^i \varphi_0 
\partial_j \varphi_0)\, ,
\end{eqnarray}
where $\nabla^{-2}$ is the inverse of the Laplacian operator. 
Therefore, inserting Eqs.~(\ref{Qf}) and~(\ref{Nf}) into 
Eq.~(\ref{traceijQN}), 
we finally obtain Eq.~(\ref{Psieq}).

Notice that, in a similar way to $N$, we can write the quantity $P$ 
defined in Eq.~(\ref{defP}) as 
\begin{equation}
\label{Pf}
P=\frac{4}{3} g^2 \left( \frac{f^2(\Omega_m)}{\Omega_m}+\frac{3}{2} \right) 
\partial^i 
\varphi_0 \partial_i \varphi_0\, .
\end{equation}

Thus applying the operator $\nabla^2 \nabla^2$ to Eq.~(\ref{PSI-PHI}) 
and combining Eq.~(\ref{Pf}) with Eq.~(\ref{Nf}) 
we find the relation between $\Psi$ and $\Phi$ of Eq.~(\ref{rela}).
\subsubsection{Vector perturbations}
Let us consider here the second-order calculation which leads to 
the generation of second-order vector and tensor modes. 
We will strictly follow the findings of Ref.~\cite{mhm}. 
For the $\Lambda=0$ case, this problem was originally solved in Ref. 
\cite{MMB} starting from the results of second-order calculations in the 
synchronous gauge \cite{tomita67,mps} and transforming them to the Poisson 
gauge, by means of second-order gauge transformations \cite{bmms}. 
In Ref.~\cite{mhm} the problem is solved directly in the Poisson 
gauge accounting for a non-vanishing 
$\Lambda$ term. 
For a non-zero-$\Lambda$ also Ref.~\cite{tomita1} studies second-order 
vector and tensor modes 
in the synchronous and Poisson gauges.

We can start by writing the second-order momentum-conservation 
equation \cite{Christoffel}
$T^\mu_{~i;\mu}=0$, which gives
\begin{equation}
\left({v}'_{2i} +{\omega}'_{2i} \right) + 
{\cal H} \left(v_{2i} + \omega_{2i} \right) =
- 2 \partial_i \Phi - 4 \dot\varphi v_{1i} 
- \partial_i \left(v_{1}^2 + \varphi^2 \right) \;.  
\end{equation}
For pure growing-mode initial conditions $\varphi' \propto \varphi$ 
and $v_{1i} \propto \partial_i \varphi$, which makes the RHS of this
equation the gradient of a scalar quantity; thus, the vector part only  
contains a decaying solution $(v^{\bot}_{2i} +\omega_{2i}) 
\propto a^{-1}$, and we can safely assume 
$v^{\bot}_{2i} = - \omega_{2i}$ (see also Ref.~\cite{MMNR} 
for a discussion on how to obtain the same results using the vorticity vector).
   
To proceed one needs the second-order perturbations of 
the Einstein tensor, $\delta_2 G^\mu_{~\nu}$: these can be found 
for any gauge in Appendix A of Refs.~\cite{ABMR,review}.  

The second-order `momentum constraint' $\delta_2 G^i_{~0} 
= 8 \pi G \delta_2 T^i_{~0}$ gives~\cite{mhm} 
\begin{equation}
\partial^i \left( {\cal H} \Phi +  \Psi' \right) 
-\frac{1}{4} \nabla^2 \omega^{2i}
+ 2 \varphi' \partial^i \varphi 
+ 8 \varphi \partial^i \varphi'
 = - 8 \pi G a^2 \bar\rho \left[\left(\varphi + 
\delta_{1}\right) v_{1}^i + v_{2}^{|| i}\right]\;. 
\end{equation}

The pure vector part of this equation can be isolated by first taking its 
divergence to solve for the combination ${\cal H}\Phi+ 
\Psi'$ and then replacing it in the original equation.
One obtains~\cite{mhm} 
\begin{equation}
- \frac{1}{2} \nabla^2 \nabla^2 \omega_{2i}  = 16 \pi G a^2 \bar\rho 
~\partial^j \left(  v_{1j} \partial_i \delta_{1} -  
v_{1i} \partial_j \delta_{1} \right)\; . 
\end{equation}

We can further simplify this equation and write: 
\begin{equation}
\label{eq:vectors}
\nabla^2  \omega_{2i} = \frac{4}{3}F(z)
\left(\partial_i\varphi_0 \nabla^2\varphi_0
- \partial^i\partial^j \varphi_0 \partial_j\varphi_0 + 
2 \partial_j\Theta_0 \right) \;, 
\end{equation}
with 
\begin{equation}
F(z)=\frac{2 g^2(z) E(z)
f(\Omega_m)} {\Omega_{0m} H_0 \left(1+z\right)^2}\;, 
\nonumber
\end{equation}
where \cite{lahav,Carroll} $f(\Omega_m) \equiv d \ln D_+/d \ln a
\approx \Omega_m(z)^{4/7}$, $H_0$ is the Hubble constant and 
\begin{equation}
\label{eq:Theta}
\nabla^2\Theta_0 = -\frac{1}{2}\left((\nabla^2\varphi_0)^2-
\partial_i \partial_k \varphi_0 \partial^i \partial^k 
\varphi_0\right) \;.
\end{equation}

For $\Lambda=0$, the above expression for $\omega_{2i}$ 
reduces to Eq. (6.8) of Ref.~\cite{MMB}, noting that $F(z)=\eta$ in
that case. 
\subsubsection{Tensor perturbations}
For the second-order tensor modes we use the traceless part of 
the $(i-j)$ Einstein equations, namely 
Eq.~(\ref{tracelessij}). To deal with this equation we replace 
it in the expression for the combination 
$\Psi-\Phi$, Eq.~(\ref{rela}), together with the expression for 
the vector mode $\omega_{2i}$. After 
some calculations one arrives at~\cite{mhm}
\begin{equation}
\label{secondgravwaves}
\nabla^2 \nabla^2
\left( {{\chi}^i_{2j}}^{\prime \prime} + 2{\cal H} 
{{\chi}^i_{2j}}^{\prime} 
- \nabla^2  \chi^i_{2j}\right)  
= 4 \nabla^2 \partial^k 
\partial_\ell R^\ell_{~k} \delta^i_j 
+ 8  \nabla^2 \left(\nabla^2 R^i_{~j} 
- \partial^i \partial_k R^k_{~j} - \partial^k \partial_j 
R^i_{~k} \right) + 4  \partial^i \partial_j \partial^k \partial_\ell 
R^\ell_{~k} 
\;, 
\end{equation}
where the traceless tensor  
\begin{eqnarray}
R^\ell_{~k} & \equiv & \partial^\ell \varphi \partial_k \varphi 
- \frac{1}{3} \left( \nabla \varphi \right)^2 \delta^\ell_k 
+ 4 \pi G a^2 \bar \rho \left( v^\ell_{1} v_{1k} - 
\frac{1}{3} v_{1}^2 \delta^\ell_k \right) \nonumber \\
& = & g^2 \left(1 + \frac{2E^2(z) f^2(\Omega_m)}{3 \Omega_{0m} 
\left(1+z\right)^3}
\right) \left( \partial^\ell \varphi_0 \partial_k  \varphi_0 
- \frac{1}{3} \left( \nabla \varphi_0 \right)^2 \delta^\ell_k \right)
\;.
\end{eqnarray}

Equation (\ref{secondgravwaves}) can be solved by Green's method, as the 
corresponding homogeneous equation is the one for linear tensor modes, whose 
analytical solutions are known in the limiting cases $\Omega_m \to 1$ or 
$\Omega_\Lambda \to 1$.
In the $\Lambda \to 0$ limit one recovers the result of Ref. \cite{MMB},
namely    
\begin{equation}
\label{eq:tensors}
\chi_{2ij}({\bf x},\eta) = \frac{1}{(2\pi)^3} \int d^3 k
e^{i{\bf k}\cdot{\bf x}}\frac{40}{k^4} {\cal S}_{ij}({\bf k})\left(
\frac{1}{3} - \frac{j_1(k\eta)}{k\eta} \right),
\end{equation}
where $j_\ell$ are spherical Bessel functions of order $\ell$ 
and ${\cal S}_{ij}({\bf k}) = \int d^3 y e^{-i{\bf k}\cdot{\bf y}}
{\cal S}_{ij}({\bf y})$, with 
\begin{equation}
\label{eq:source}
{\cal S}_{ij}=\nabla^2 \Theta_0 \delta_{ij} + \partial_i 
\partial_j \Theta_0 + 2\left( \partial_i \partial_j 
\varphi_0 \nabla^2\varphi_0 - \partial_i \partial_k
\varphi_0 \partial^k \partial_j \varphi_0\right) \;. 
\end{equation}

\vskip 1cm
\setcounter{equation}{0}
\def\theequation{B.\arabic{equation}}
\vskip 0.2cm
\section{Derivation of the multipoles $\mathop{a^{\rm NL}_{\ell m}}$}
\label{B}
In this section we will show how to obtain Eq.~(\ref{almNL}). 
We plug Eq.~(\ref{ISW2nd}) into 
Eq.~(\ref{alm}).  In Eq.~(\ref{ISW2nd}) the 
two integrals depend on the time derivative of the second-order 
gravitational potentials $(\Phi+\Psi)'$, while 
the contribution ${\rm (first-order)^2}_{\rm late}$ comes from additional 
second-order corrections. 
For simplicity let us compute these two parts separately in Eq~(\ref{ISW2nd})
\begin{equation}
\frac{1}{2} {\frac{\Delta T_2}{T}}^{\rm late}(\hat{\bf n})=\frac{1}{2} 
\frac{\Delta T_2}{T}[\Phi'+\Psi']+({\rm first-order})_{\rm late}^2
\, ,
\end{equation}
where
\begin{eqnarray}
\label{DTPSI}
\frac{1}{2}\frac{\Delta T_2}{T}[\Phi'+\Psi'] &=&
\int_{\eta_m}^{\eta_0} d\eta\, 
\Bigg[ \left( -\frac{10}{3}(a_{\rm nl}-1) g_{m} 
g'(\eta)-2 g_{m} g'(\eta) + B'_1(\eta)
+ 4 g(\eta) g'(\eta)
\right) \varphi_0^2  
\nonumber \\
&+& g^2_{m} \overline{B}(\eta)
\left( \nabla^{-2} 
\left( \partial^i \varphi_0 
\partial_i \varphi_0 \right)- 3 \nabla^{-4} \partial_i \partial^j
\left(\partial^i \varphi_0 \partial_j \varphi_0 \right) \right)
\Bigg] \Bigg|_{{\bf x}=- \hat{\bf n}(\eta_0-\eta)} \, ;
\end{eqnarray}
here $\overline{B}(\eta)$ is given in Eq.~(\ref{barB}) and 
\begin{eqnarray}
\label{correxpl}
({\rm first-order})_{\rm late}^2&=&-4 
\int_{\eta_m}^{\eta_0} d\eta\, \Bigg[  
g(\eta) g'(\eta) \varphi^2_0(\eta,{\bf x}) + 
\left[  \int_{\eta_0}^\eta d\tilde{\eta} \varphi \right] 
g''(\eta) \varphi_0(\eta,{\bf x}) \Bigg] 
\Bigg|_{{\bf x}=- \hat{\bf n}(\eta_0-\eta)} \nonumber \\
&+&\frac{1}{2} I^2_1(\eta_m) -
\frac{1}{3} \varphi_*I_1(\eta_m)\, ,
\end{eqnarray}
where we have used Eq.~(\ref{relphiphi_0}) into Eq.~(\ref{corrf}).  
It is important to notice that in Eq.~(\ref{correxpl}) 
$\varphi_*\equiv 
\varphi(\eta_*,{\bf x}=-{\hat{\bf n}}(\eta_0-\eta_*))$ and the 
gravitational potential under a given integral must be evaluated 
along the background geodesics, 
so that, for example,  
\begin{equation}
\int_{\eta_0}^\eta d\tilde{\eta}\, \varphi= \int_{\eta_0}^{\eta} 
d \tilde{\eta}\, \varphi
\Big |_{{\bf x}=- \hat{\bf n}(\eta_0-\tilde{\eta})}=
\int_{\eta_0}^{\eta} d \tilde{\eta}\, g'(\tilde{\eta}) 
\varphi_0(\tilde{\eta},{\bf x})  
\Big |_{{\bf x}=- \hat{\bf n}(\eta_0-\tilde{\eta})}\, .
\end{equation}
Correspondingly in Eq.~(\ref{alm}) we will have 
\begin{equation}
a^{\rm NL}_{lm}=a^{\rm NL}_{lm}[\Phi'+\Psi']+
a^{\rm NL}_{lm}\left[{\rm (first-order)^2}\right]
\, .
\end{equation}  
Let us consider the term in Eq.~(\ref{DTPSI}). We express it 
in terms of its Fourier transform
\begin{eqnarray}
\label{FDTPSI}
\frac{1}{2}\frac{\Delta T_2}{T}[\Phi'+\Psi'] &=&
\int \frac{d^3 k}{(2 \pi)^3} \int_{\eta_m}^{\eta_0} d\eta\, 
\Bigg[ \left( -\frac{10}{3}(a_{\rm nl}-1) g_{m} g'(\eta)-2 
g_{m} g'(\eta) + B'_1(\eta)
+ 4 g(\eta) g'(\eta)
\right) [\varphi_0^2]({\bf k}) 
\nonumber \\
&-& g_{m}^2 
\overline{B}(\eta) K_2({\bf k}) \Bigg]   e^{- i {\bf k}\cdot 
{\hat{\bf n}}(\eta_0-\eta)}\, ,
\end{eqnarray}  
where $K_2({\bf k})$ is defined in Eq.~(\ref{def:Kn}) and 
$[\varphi_0^2]({\bf k})$ 
is the convolution giving the Fourier transform of $\varphi^2_0({\bf x})$. 
Inserting Eq.~(\ref{FDTPSI}) into Eq.~(\ref{alm}) and using the Rayleigh 
expansion 
\begin{equation}
\label{Rayleigh}
e^{- i {\bf k} \cdot {\bf x}}=4 \pi
\sum_{\ell m} (-i)^\ell\, j_\ell(kx) Y^*_{\ell m}(\hat{\bf k}) 
Y_{\ell m}(\hat{\bf x})
\end{equation}
with the orthonormality of the spherical harmonics  we find 
\begin{eqnarray}
\label{almNLPSI}
a^{\rm NL}_{lm}[\Phi'+\Psi']&=& 
4 \pi(-i)^\ell \int \frac{d^3 k}{(2\pi)^3} 
\int_{\eta_m}^{\eta_0} d\eta\, 
j_\ell(k(\eta_0-\eta))\, \Bigg[ \left( -\frac{10}{3}(a_{\rm nl}-1)  
\frac{g'(\eta)}{g_{m}}-2 
\frac{g'(\eta)}{g_{m}} + \frac{B'_1(\eta)}{g_{m}^2}
+ 4 \frac{g(\eta) g'(\eta)}{g_{m}^2}
\right) [\varphi_0^2]({\bf k}) 
\nonumber \\
&-& \overline{B}(\eta) K_2({\bf k}) \Bigg] Y^*_{lm}(\hat{\bf k})\, ,
\end{eqnarray} 
where we have used Eq.~(\ref{relphiphi_0}).

Let us now consider the term in Eq.~(\ref{correxpl}). We Fourier expand 
the gravitational potential 
under each integral and we use the Rayleigh expansion~(\ref{Rayleigh}). 
For the first term in 
Eq.~(\ref{correxpl}) the computation is the same as described above, 
while the other terms has a 
different dependence on $\hat{\bf n}$. For instance, in   
Eq.~(\ref{correxpl}) the integral over $g''(\eta)$ becomes
\begin{eqnarray}
\label{example}  
&-&4 \int_{\eta_m}^{\eta_0} d\eta\, 
\left[  \int_{\eta_0}^\eta d\tilde{\eta} \varphi \right] 
g''(\eta) \varphi_0(\eta,{\bf x})
\Bigg|_{{\bf x}=- \hat{\bf n}(\eta_0-\eta)}=
\nonumber \\
&=& -4 \int \frac{d^3k_1}{(2 \pi)^3} \frac{d^3k_2}{(2 \pi)^3} 
\int_{\eta_m}^{\eta_0} d\eta \frac{g''(\eta)}{g_{m}}
\int_{\eta_0}^{\eta} d\tilde{\eta} \frac{g(\tilde{\eta})}{g_{m}}   
\varphi_{m}({\bf k}_1) \varphi_{m}({\bf k}_2) 
e^{- i {\bf k}_1\cdot {\hat{\bf n}}(\eta_0-\eta)}
e^{- i {\bf k}_2\cdot {\hat{\bf n}}(\eta_0-\tilde{\eta})} \nonumber \\
&=& -4\,
(4 \pi)^2 \sum_{L_1M_1} \sum_{L_2M_2} (-i)^{L_1+L_2}\, 
Y^*_{L_1M_1}(\hat{\bf n}) 
Y^*_{L_2M_2}(\hat{\bf n})  
\int \frac{d^3k_1}{(2 \pi)^3} \frac{d^3k_2}{(2 \pi)^3} 
\int_{\eta_m}^{\eta_0} d\eta \frac{g''(\eta)}{g_{m}} 
j_{L_1}(k_1(\eta_0-\tilde{\eta})) \nonumber \\
& & \int_{\eta_0}^\eta d\tilde{\eta} \frac{g(\tilde{\eta})}{g_{m}} 
j_{L_2}(k_2(\eta_0-\tilde{\eta})) 
\varphi_{m}({\bf k}_1) 
\varphi_{m}({\bf k}_2) Y_{L_1M_1}(\hat{\bf k}_1) 
Y_{L_2M_2}(\hat{\bf k}_2)\, , 
\nonumber \\
\end{eqnarray}  
and similar expressions hold for all of the three remaining terms 
in Eq.~(\ref{correxpl}). 
Eq.~(\ref{example}) shows that in general 
$\Delta T/T$ does not depend on $\hat{\bf n}$ only through a single angle 
$\bf k \cdot \hat{\bf n}$, as it happens in linear theory.

Thus Eq.~(\ref{correxpl}) becomes
\begin{eqnarray}
\label{FOS}
& & ({\rm first-order})_{\rm late}^2=
- 4 \int \frac{d^3 k}{(2 \pi)^3} \int_{\eta_*}^{\eta_0} d\eta\, 
g(\eta) g'(\eta) 
[\varphi^2_{m}]({\bf k}) e^{- i {\bf k}\cdot 
{\hat{\bf n}}(\eta_0-\eta)} \nonumber \\
& & + (4 \pi)^2 \sum_{L_1M_1} \sum_{L_2M_2} (-i)^{L_1+L_2}\,
 Y^*_{L_1M_1}(\hat{\bf n}) 
Y^*_{L_2M_2}(\hat{\bf n}) 
\int \frac{d^3k_1}{(2 \pi)^3} \frac{d^3k_2}{(2 \pi)^3} 
\varphi_{m}({\bf k}_1) 
\varphi_{m}({\bf k}_2) \Delta_{L_1L_2}(k_1,k_2) Y_{L_1M_1}(\hat{\bf k}_1) 
Y_{L_2M_2}(\hat{\bf k}_2)
\, ,
\nonumber \\
\end{eqnarray} 
where $\Delta_{L_1L_2}(k_1,k_2)$ is the function defined in Eq.~(\ref{TR}). 
From Eq.~(\ref{FOS}) it is straightforward to compute the 
corresponding multipoles using Eq.~(\ref{alm}). We find
\begin{eqnarray}
\label{almNLFOS}
a^{\rm NL}_{lm}\left[{\rm (first-order)^2}\right]&=& - 
4 \pi(-i)^\ell \int \frac{d^3 k}{(2\pi)^3} 
\int_{\eta_*}^{\eta_0} d\eta\, 
j_\ell(k(\eta_0-\eta))\,  
\frac{g(\eta) g'(\eta)}{g_{m}^2}
\, [\varphi_0^2]({\bf k})\,  Y^*_{\ell m}(\hat{\bf k}) \nonumber \\
&+&
(4\pi)^2 \sum_{L_1 M_1} \sum_{L_2 M_2} (-i)^{L_1+L_2} 
{\mathcal G}^{m M_1 L_1}_{\ell L_1 L_2}
\int \frac{d^3 k_1}{(2\pi)^3}  \frac{d^3 k_2}{(2\pi)^3} 
\varphi_{m}({\bf k}_1) 
\varphi_{m}({\bf k}_2) \nonumber \\
&\times& \Delta_{L_1 L_2}(k_1,k_2) Y_{L_1 M_1}(\hat{\bf k}_1) 
Y_{L_2 M_2}(\hat{\bf k}_2)\, . \nonumber \\
\end{eqnarray}
The sum of Eq.~(\ref{almNLPSI}) and Eq.~(\ref{almNLFOS}) 
gives the non-Gaussian part of the 
multipoles $a^{\rm NL}_{\ell m}$ for the late ISW effect written in Eq.~(\ref{almNL}).    

\vskip 1cm
\setcounter{equation}{0}
\def\theequation{C.\arabic{equation}}
\vskip 0.2cm
\section{Calculation of the Early Integrated Sachs-Wolfe effect at second order}
\label{C}
We consider a flat Universe filled by radiation $(\rho_\gamma)$ and pressureless 
matter $(\rho_m)$. In this Appendix we derive the equations~(\ref{relPsiPhimr}) and~(\ref{eqPSI}). 

The relation~(\ref{relPsiPhimr}) is obtained through 
the same steps of Section $3$ in Appendix~\ref{A}. 
We take the traceless part of the ($i$-$j$)-component of Einstein equations 
at second-order 
$\delta_2 G^i_{~j} - \frac{1}{3}\delta_2G^k_{~k} 
\delta^i_j = 8 \pi G (\delta_2T^i_{~j} - 
\frac{1}{3} \delta_2T^k_{~k} \delta^i_j)$.
The second-order perturbations of the Einstein tensor $G^\mu_{~\nu}$ 
can be found for any gauge in 
Appendix A of Refs.~\cite{ABMR,review}. For the total energy momentum tensor 
$T_{\mu \nu}= T^{(\gamma)}_{\mu \nu}+T^{(m)}_{\mu \nu}$ one finds 
\begin{eqnarray}
\delta_2T^i_{~j}= \left( 
\frac{w_{\gamma}}{2} \rho_\gamma 
\delta_{2\gamma} +\frac{w_m}{2} \rho_m \delta_{2m}\right) \delta^i_{~j} +
\rho_\gamma (1+w_\gamma) v^i_{1\gamma} v^1_{j\gamma}+ \rho_m (1+w_m)  
v^i_{1m} v^1_{jm}\, ,
\end{eqnarray}
where $w_\gamma=1/3$ and $w_m=0$ are the equations of state of radiation and matter respectively, 
$v^i_{1\gamma}$, $v^i_{1m}$ their linear (scalar) velocities and $\delta_{2\gamma}=
\delta_2 \rho_\gamma/\rho_\gamma$.   Thus we obtain 
\begin{eqnarray}
\label{tracelessijmr}
&& - \left[\frac{1}{6}\nabla^2\left(\Psi - \Phi \right) + 
\frac{2}{3} \left(\nabla \varphi\right)^2 + 
\frac{4}{3} \varphi \nabla^2 \varphi \right] 
\delta^i_j + \frac{1}{2} \partial^i \partial_j 
\left(\Psi - \Phi \right) 
+ 2 \partial^i \varphi \partial_j \varphi + 4 \varphi \partial^i \partial_j 
\varphi \nonumber \\
&& 
- \frac{1}{4} \left(\partial^i {{\omega}_{2j}}{\prime}  + \partial_j 
{{\omega}_2^i}^{\prime} \right) - \frac{1}{2} {\cal H} 
\left( \partial^i \omega_{2j}+ \partial_j 
\omega_2^i \right) +
\frac{1}{4} \left( {{\chi}^i_{2j}}^{\prime \prime} + 2{\cal H} 
{{\chi}^i_{2j}}^{\prime} 
- \nabla^2  \chi^i_{2j}\right)  = \nonumber \\
& & 
\frac{3{\cal H}^2}{\rho} \left[ 
\rho_\gamma (1+w_\gamma) v^i_{1\gamma} v^1_{j\gamma}-\frac{1}{3} \rho_\gamma (1+w_\gamma) 
v^2_{1\gamma} 
+\rho_m (1+w_m) v^i_{1m} v^1_{jm}-\frac{1}{3} \rho_m (1+w_m) 
v^2_{1m} \right]\, ,
\end{eqnarray}
where $\rho=\rho_\gamma+\rho_m$ and 
$v^2_{1\gamma}=v^i_{1\gamma} v^1_{i\gamma}$. We now apply the operator 
$\partial_i \partial^j$ to get rid of the vector and tensor modes, and to 
solve for the combination $(\Psi-\Phi)$ 
\begin{equation}
\label{PSI-PHImr}
\Psi-\Phi=-4\varphi^2+{\cal Q}\, ,
\end{equation}
where ${\cal Q}$ is defined by 
\begin{equation}
\label{defQmr}
\nabla^2{\cal Q} = -P+3 N\, ,
\end{equation}
with 
\begin{equation}
\label{defPmr}
P \equiv P^i_{~i}\, ,
\end{equation}
where
\begin{equation}
\label{defPijmr}
P^i_{~j} = 2 \partial^i \varphi \partial_j \varphi + 3 {\cal H}^2(1+w_\gamma)\, 
\frac{\rho_\gamma}{\rho}\, 
v^i_{1\gamma}v^1_{j\gamma}+ 3 {\cal H}^2(1+w_m)\, \frac{\rho_m}{\rho}\,  
v^i_{1m}v^1_{jm} \, ,
\end{equation}
and the quantity $N$ is given by 
\begin{equation}
\label{defNmr}
\nabla^2 N \equiv \partial_i \partial^j P^i_j\, .
\end{equation}

In order to get the explicit expression for ${\cal Q}$, Eq.~(\ref{exprQ}), we need to know the 
linear velocities $v^1_{j\gamma}$ and $v^1_{jm}$. For adiabatic initial conditions on large scales 
(deep in the radiation dominated era) one has that initially  $v^1_{j\gamma}=v^1_{jm}$ (see, {\it 
e.g.}, Ref.~\cite{mabert}). Moreover it can be shown that on large-scales for adiabatic 
perturbations the velocities then remain equal (see, {\it e.g.} Refs.~\cite{mw2,kodamasasaki}). 
Therefore we set 
$v^1_{j\gamma}=v^1_{jm}\equiv v^1_j$ and from the momentum constraint ($0$-$i$) 
Einstein equation we 
have the expression of the velocities in term of $\varphi$
\begin{equation}
\label{vphi}
\partial_i(\varphi'+{\cal H} \varphi)=-\frac{3}{2} {\cal H}^2 \left( 
\frac{4}{3} \frac{\rho_\gamma}{\rho}\, v^1_{i\gamma}+ 
\frac{\rho_m}{\rho}v^1_{im}  \right) = 
-\frac{3}{2} {\cal H}^2  \left(1+\frac{1}{3} \frac{\rho_\gamma}{\rho} \right)\, v^1_i\, ,
\end{equation}
where we have used that $w_\gamma=1/3$ abd $w_m=0$. 

Using Eq.~(\ref{vphi}) in Eq.~(\ref{defPijmr}) brings
\begin{equation}
\label{Pijmr}
P^i_{~j}= \frac{1}{2\pi G a^2 \rho} \left( 1+\frac{1}{3} \frac{\rho_\gamma}{\rho} \right)^{-1} 
\left[ \partial^i \varphi' \partial_j \varphi'+{\cal H} (
\partial^i \varphi \partial_j \varphi'+\partial^i \varphi' \partial_j \varphi) 
+\frac{{\cal H}^2}{2}\left( 5+\frac{\rho_\gamma}{\rho} \right) \partial^i\varphi 
\partial_j \varphi \right]\, .
\end{equation}
We now make use of the evolution for the linear growing mode of the gravitational potential for 
adiabatic perturbations~\cite{HuEis} 
\begin{equation}
\varphi=- F(\eta) \zeta_1\, ,
\end{equation} 
where $\zeta_1$ is the curvature perturbation which is a constant on large scales 
for adiabatic perturbations, $\zeta_1={\rm const}$. Equivalently we can use 
Eq.~(\ref{varphimr}), with 
$\zeta_1=- \varphi_*/F_*$. The function $F(\eta)$ is given in Eq.~(\ref{defF}).

Using Eq.~(\ref{Pijmr}) we thus find
\begin{eqnarray}
\nabla^2 \nabla^2 N &=& \frac{4}{3} \left(1+\frac{\rho_\gamma}{\rho}\right)^{-1}
\frac{1}{F_*^2} \left[ \left( \frac{F'}{{\cal H}}\right)^2+2F\frac{F'}{\cal H}+\frac{1}{2} 
\left( 5+\frac{\rho_\gamma}{\rho} \right) F^2 \right] \nabla^2 \partial_i\partial^j (
\partial^i \varphi_* \partial_j\varphi_*)\, , \\
P&=&\frac{4}{3} \left(1+\frac{\rho_\gamma}{\rho}\right)^{-1}
\frac{1}{F_*^2} \left[ \left( \frac{F'}{{\cal H}}\right)^2+2F\frac{F'}{\cal H}+\frac{1}{2} 
\left( 5+\frac{\rho_\gamma}{\rho} \right) F^2 \right] 
\partial^i \varphi_* \partial_i\varphi_*\, ,
\end{eqnarray}
and from Eq.~(\ref{defQmr})
\begin{equation}
{\cal Q}= -\frac{4}{3}\left(1+\frac{\rho_\gamma}{\rho}\right)^{-1}
\frac{1}{F_*^2} \left[ \left( \frac{F'}{{\cal H}}\right)^2+2F\frac{F'}{\cal H}+\frac{1}{2} 
\left( 5+\frac{\rho_\gamma}{\rho} \right) F^2 \right] \left[ \nabla^{-2} (\partial^i \varphi_* 
\partial_i \varphi_*)-3 \nabla^{-4} \partial_i\partial^j (\partial^i \varphi_* 
\partial_j \varphi_*)    \right] \, .
\end{equation}
Notice that the second squared bracket is $3{\cal K}/10$, where ${\cal K}$ is given by 
Eq.~(\ref{calK}). Finally by introducing the ratio $y=\rho_m/\rho_\gamma$ and using $d/d\ln a=
{\cal H}^{-1} d/d\eta$ we can write 
\begin{equation} 
{\mathcal Q} = - \frac{3}{10 F_*^2}\, \widetilde{Q}\, {\mathcal K}\, ,
\end{equation} 
where ${\widetilde Q}$ is given by Eq.~(\ref{tildeQ}).

We now describe in some details how to obtain the equation~(\ref{eqPSI}) 
for the gravitational potential $\Psi$. 
We start from the expression for the curvature perturbation $\zeta_2$ in the Poisson gauge, 
see Eq.~(\ref{defz2}), 
\begin{equation}
\label{defz2P}
\zeta_2=-\Psi-{\cal H} \frac{\delta_2 \rho}{\rho'}+\Delta \zeta_2\, ,
\end{equation}
and the ($0$-$0$) Einstein equation on large scales 
\begin{equation}
3{\cal H}(\Psi'+{\cal H} \Phi)=8 \pi G \delta_2 T^0_{~0}+12{\cal H}^2 \varphi^2 +3 
{\varphi'}^{2}\, .
\end{equation}
In Eq.~(\ref{defz2P}) the second-order corrections $\Delta \zeta_2$ are given by 
\begin{equation}
\label{deltaz2P}
\Delta \zeta_2 = 2 {\cal H} \frac{\delta_1 \rho'}{\rho'} \frac{\delta_1 \rho}{\rho'}+2
\frac{\delta_1 \rho}{\rho'} (\varphi'+2{\cal H} \varphi) - 
\left( \frac{\delta_1 \rho}{\rho'} \right)^2 \left({\cal H} \frac{\rho''}{\rho} -{\cal H}' 
-2{\cal H}^2  \right)\, ,
\end{equation}
while the second-order perturbations of the Einstein tensor $G^\mu_{~\nu}$ can be found 
for any gauge in Appendix A of Refs.~\cite{ABMR,review}. The ($0$-$0$) component of the energy 
momentum tensor reads
\begin{equation}
\label{T002}
\delta_2 T^0_{~0}=-\frac{1}{2}\delta_2 \rho-\rho_m v^2_{1m}-\frac{4}{3} \rho_\gamma v^2_{1\gamma}\, ,
\end{equation} 
where $\rho$ is the total energy density. We thus express $\delta_2 \rho$ in terms of the 
gravitational potential through the ($0$-$0$) Einstein equation, and we insert it in  
Eq.~(\ref{T002}) to find
\begin{equation}
\label{rhoz}
\zeta_2=-\Psi+2\frac{\rho}{\rho'}(\Psi'+{\cal H}\Phi)-8\frac{\rho}{\rho'}{\cal H} \varphi^2 
-2 \frac{\rho}{{\cal H} \rho'} {\varphi'}^{2}+\Delta \zeta_2\, ,
\end{equation}
where we have dropped the velocities appearing in Eq.~(\ref{deltaz2P}) since they are negligible 
on large scales. Finally we use the relation~(\ref{PSI-PHImr}) found previously 
between the two gravitational potentials to eliminate $\Phi$ in Eq.~(\ref{rhoz})
\begin{equation}
\label{zPSI}
\zeta_2=-\Psi+2\frac{\rho}{\rho'}(\Psi'+{\cal H}\Psi)-2\frac{\rho}{\rho'} {\cal H} {\cal Q} 
-2 \frac{\rho}{{\cal H} \rho'} {\varphi'}^{2}+\Delta \zeta_2\, .
\end{equation}

Switching time variable from $d/d\eta$ to $^{\displaystyle{\cdot}} \equiv d/d\ln a 
= {\cal H}^{-1} d/d\eta$ we notice that 
\begin{equation}
-\frac{1}{2} \frac{\dot{\rho}}{\rho} \Psi+(\dot{\Psi}+\Psi) = \frac{\sqrt \rho}{a} \left[ \frac{a}{\sqrt \rho}
 \Psi \right]^{\displaystyle{\cdot}}\, ,
\end{equation}
and thus Eq.~(\ref{zPSI}) becomes
\begin{equation}
\label{eqPSIappendix}
\frac{\sqrt \rho}{a} \left[ \frac{a}{\sqrt \rho} \Psi \right]^{\displaystyle{\cdot}}=\frac{1}{2} 
\frac{\dot{\rho}}{\rho} \zeta_2+ \left[ \dot{\varphi}^2+{\mathcal Q}
-\frac{1}{2}\frac{\dot{\rho}}{\rho}  \Delta \zeta_2\right]\, .
\end{equation}

In the rest of this Appendix we compute the expression~(\ref{exprDz2}) for the quantity $\Delta \zeta_2$ 
defined in 
Eq.~(\ref{deltaz2P}). This is obtained by evaluating $\Delta \zeta_2$ in the case of radiation plus 
pressureless 
matter. 

First of all from the the energy continuity equations $\rho_\gamma'=-4 {\cal H} \rho_\gamma$ and 
$\rho_m'=-3 {\cal H} \rho_m$ one obtains for the total energy density $\rho=\rho_\gamma+\rho_m$ 
\begin{equation}
\frac{\rho''}{\rho'}=\frac{{\cal H}'}{\cal H}-\frac{9\rho_m+16\rho_\gamma}{3\rho_m+4\rho_\gamma}\, ,
\end{equation}
and hence 
\begin{equation}
\label{comb}
{\cal H} \frac{\rho''}{\rho} -{\cal H}' -2{\cal H}^2 =-{\cal H}^2 
\left( 6-\frac{3\rho_m}{3\rho_m+4\rho_\gamma}\right)\, ,
\end{equation}
which is the combination appearing in Eq.~(\ref{deltaz2P}). 

We can now express $\delta_1\rho$ and $\delta_1\rho'$ in terms of the graviational potential $\varphi$ using the 
total energy continuity equation and the ($0$-$0$) Einstein equation in the large scale limit. 
The total energy continuity equation on large scales reads
\begin{equation}
\delta_1\rho'=-3 {\cal H}(\delta_1\rho +\delta_1 p)-3(\rho+p)\varphi'=0\, ,
\end{equation}
where $p=p_\gamma+p_m=\rho_\gamma/3$, and $\delta_1 \rho=\delta_1 \rho_\gamma+\delta_1 \rho_m$, 
$\delta_1 p =\delta_1 p_\gamma+\delta_1 p_m=\delta_1 \rho_\gamma/3$ are the total energy density and pressure 
perturbation respectively. We now take advantage of the adiabaticity of the perturbations 
\begin{equation}
\frac{\delta_1\rho_\gamma}{\rho_\gamma}=\frac{4}{3}\frac{\delta_1\rho_m}{\rho_m}\, ,
\end{equation}
to find from the continuity equation
\begin{equation}
\label{drho'}
\frac{\delta_1\rho'}{\rho'}=-3{\cal H}\left(1+\frac{1}{3} \frac{4\rho_\gamma}{4\rho_\gamma+3\rho_m} \right)
\frac{\delta_1 \rho}{\rho'}-\frac{\varphi'}{\cal H}\, .
\end{equation}
Plugging Eqs.~(\ref{comb}) and~(\ref{drho'}) into Eq.~(\ref{deltaz2P}) yields
\begin{equation}
\Delta \zeta_2=-{\cal H}^2 \left( \frac{\delta_1 \rho}{\rho'} \right)^2 \left(1+ 
\frac{4\rho_\gamma}{4\rho_\gamma+3\rho_m} \right)
+4{\cal H}\varphi \frac{\delta_1 \rho}{\rho'} \, .
\end{equation}
Finally we use the first-order ($0$-$0$) Einstein equation on large scales
\begin{equation}
\varphi'+{\cal H}\varphi=-\frac{1}{2} {\cal H}\frac{\delta_1 \rho}{\rho}\, ,
\end{equation}
to get 
\begin{equation}
\Delta \zeta_2=\left[ 2\frac{\rho}{\dot{\rho}} (\dot{\varphi}+\varphi) \right]^2 
\left(1+\frac{1}{3} \frac{4\rho_\gamma}{4\rho_\gamma+3\rho_m} \right)-8\frac{\rho}{\dot{\rho}} 
\varphi (\dot{\varphi}+\varphi)\, ,
\end{equation}
where a dot stands by $d/d \ln a$.

We now use Eq.~(\ref{varphimr}), which gives linear growing mode for the gravitational potential, to find
\begin{eqnarray}
-\frac{1}{2}\frac{\dot{\rho}}{\rho} \Delta \zeta_2&=&\frac{1}{F_*^2} 
\left[ -2\left( \frac{1+y}{4+3y} \right)\left( 1+\frac{4}{4+3y} \right) \dot{F}^2+
2\left( 2-\left( \frac{1+y}{4+3y} \right)\left( 1+\frac{4}{4+3y} \right) \right)F^2
\right. \nonumber \\
&+& \left. 4\left( 1-\left( \frac{1+y}{4+3y} \right)\left( 1+\frac{4}{4+3y} \right) \right) F\dot{F} 
\right] \varphi_*^2\, ,
\end{eqnarray}
where $y=\rho_m/\rho_\gamma$. With the definition
\begin{equation}
R(y)=\frac{1+y}{4+3y}\left( 1+\frac{4}{4+3y} \right)\, ,
\end{equation}
we get Eq.~(\ref{exprDz2}).



\begin{references}
\bibitem{Christoffel} 
Second-order Christoffel symbols can be found in 
Appendix A of Ref.~\cite{ABMR,review} (see also Ref.~\cite{noh})
\end{references}
\end{document}